\documentclass[12pt]{article}

\usepackage[dvipsnames]{xcolor}
\usepackage{etex}
\usepackage[bulletsep]{collref}
\usepackage{amssymb,graphicx}
\usepackage[intlimits]{amsmath}
\usepackage{pst-all}
\usepackage{bbm}
\usepackage[small]{subfigure}
\usepackage{pstricks,pst-node}

\usepackage{MnSymbol}

\makeatletter \@addtoreset{equation}{section} \makeatother

\makeatletter
\let\old@startsection=\@startsection
\let\oldl@section=\l@section
\renewcommand{\@startsection}[6]{\old@startsection{#1}{#2}{#3}{#4}{#5}{#6\mathversion{bold}}}
\renewcommand{\l@section}[2]{\oldl@section{\mathversion{bold}#1}{#2}}
\makeatother

\makeatletter
\let\old@makecaption=\@makecaption
\def\@makecaption{\small\old@makecaption}
\makeatother

\newcommand{\pictext}[1]{
     \raisebox{-4.5ex}{\includegraphics[height=10ex]{#1}}
}

\newcommand{\bea}{\begin{equation}}
\newcommand{\eea}{\end{equation}}
\newcommand{\bear}{\begin{eqnarray}}
\newcommand{\eear}{\end{eqnarray}}
\newcommand{\bearr}{\begin{eqnarray*}}
\newcommand{\eearr}{\end{eqnarray*}}

\begin{document}

\begin{flushright}\footnotesize
\texttt{NORDITA-2012-49} \\
\texttt{UUITP-18/12}
\vspace{0.6cm}
\end{flushright}

\renewcommand{\thefootnote}{\fnsymbol{footnote}}
\setcounter{footnote}{0}

\begin{center}
{\Large\textbf{\mathversion{bold} Ladders for Wilson Loops  \\
Beyond Leading Order}
\par}

\vspace{0.8cm}

\textrm{D.~Bykov$^{1,2}$ and
K.~Zarembo$^{1,3}$\footnote{Also at ITEP, Moscow, Russia}}
\vspace{4mm}

\textit{${}^1$Nordita,
Roslagstullsbacken 23, SE-106 91 Stockholm, Sweden}\\
\textit{${}^2$Steklov Mathematical Institute of Russ. Acad. Sci., Gubkina str. 8, 119991 Moscow, Russia}\\
\textit{${}^3$Department of Physics and Astronomy, Uppsala University\\
SE-751 08 Uppsala, Sweden}\\
\vspace{0.2cm}
\texttt{dbykov@nordita.org, zarembo@nordita.org}

\vspace{3mm}


\par\vspace{1cm}

\textbf{Abstract} \vspace{3mm}

\begin{minipage}{13cm}
We set up a general scheme to resum ladder diagrams for the quark-anti-quark potential in $\mathcal{N}=4$ super-Yang-Mills theory,  and do explicit calculations at the next-to-leading order. The results perfectly agree with string theory in $AdS_5\times S^5$ when continued to strong coupling, in spite of a potential order-of-limits problem.
\end{minipage}

\end{center}

\vspace{0.5cm}


\newpage
\setcounter{page}{1}
\renewcommand{\thefootnote}{\arabic{footnote}}
\setcounter{footnote}{0}

\section{Introduction}
 
The potential between static charges in gauge theories is defined through the expectation value of a rectangular Wilson loop. In the AdS/CFT context, Wilson loops are directly related to string worldsheets and at strong coupling their expectation values obey the area law \cite{Maldacena:1998im,Rey:1998ik,Drukker:1999zq}. The static potential in $\mathcal{N}=4$ super-Yang-Mills theory (SYM) can thus be computed at strong coupling by evaluating the area of a minimal surface in $AdS_5\times S^5$ with appropriate boundary conditions \cite{Maldacena:1998im,Rey:1998ik}. This calculation has led to one of the first quantitative predictions of the AdS/CFT duality \cite{Maldacena:1998im}. Recently much progress has been achieved towards a non-perturbative description of the static potential at any coupling. The main tools are diagram resummations \cite{Correa:2012nk} and quantum integrability of the AdS/CFT system \cite{Drukker:2012de,Correa:2012hh}. The former approach is less general and is applicable only in a corner of the parameter space,  but is technically simpler and can be derived from first principles. Apart from giving another insight into the inner working of the AdS/CFT duality, this approach can be helpful in finding direct gauge-theory interpretation of various quantities appearing in TBA (Y-functions, driving terms in the TBA equations {\it etc.}).

The approach of \cite{Correa:2012nk} is based on the resummation of  ladder diagrams. The Bethe-Salpeter equation that resums ladders for rectangular Wilson loop was derived in \cite{Erickson:1999qv} and predicts the $\sqrt{\lambda }$ scaling of the static potential, consistent with the string calculation at strong coupling. The limit that emphasizes ladder diagrams involves analytic continuation in the scalar coupling of the Wilson loop \cite{Correa:2012nk}, and once the same limit is taken on the two sides of the duality, the results completely agree \cite{Correa:2012nk}, in spite of a potential order-of-limits problem. Identification of the ladder approximation with the leading order of an expansion in a small parameter \cite{Correa:2012nk} makes it possible to systematically compute corrections\footnote{See \cite{Shuryak:fk} for an early discussion of how to improve the ladder approximation.}. Our aim will be to set up a general scheme of doing ladder resummation order by order. We will explicitly compute the leading correction to the simple ladder approximation. In particular we will check if the order-of-limits problem start to affect the comparison to string theory at strong coupling, once the NLO corrections are taken into account.
 
\section{Coulomb potential} 

The Wilson loop operator in $\mathcal{N}=4$ SYM theory is defined as
\begin{equation}
 W(C)=\mathop{\mathrm{tr}}{\rm P}\exp\left[\oint_C ds\left(A_\mu \dot{x}^\mu +i\Phi _In^I|\dot{x}|\right)
 \right]
\end{equation}
It describes an infinitely heavy W-boson associated with the symmetry breaking by a Higgs vev in the direction of the six-dimensional unit vector $\mathbf{n}$: $\left\langle \Phi _I\right\rangle=vn_I$. Rectangular $T\times L$ loop with the scalar couplings $\mathbf{n}_1$ and $\mathbf{n}_2$ on the two  sides of the rectangle defines the Coulomb potential between $W$ and $\bar{W}$ with the symmetry-breaking vevs $\mathbf{n}_{1,2}$:
\begin{equation}\label{logW}
 \ln \left\langle W(C_{T\times L,\mathbf{n}_1,\mathbf{n}_2})\right\rangle
=\frac{\alpha (\theta ,\lambda )T}{L}\,.
\end{equation}
The Coulomb charge $\alpha (\theta ,\lambda )$ is a function of the 't~Hooft coupling $\lambda =g^2_{\rm YM}N$ and the angle between the symmetry-breaking vevs: $\cos\theta =\mathbf{n}_1\cdot \mathbf{n}_2$.

To the leading order in  perturbation theory,
\begin{equation}
 \alpha (\theta ,\lambda )=\frac{\lambda }{8\pi }\left(1+\cos\theta \right)+O\left(\lambda ^2\right),
\end{equation}
where the first term in the brackets comes from the gluon exchange and the second term from exchanging scalars. The limit, which emphasizes the ladder diagrams, requires an analytic continuation in $\theta $:
\begin{equation}\label{vartheta}
 \theta =i\vartheta.
\end{equation}
When $\vartheta$ is real and large, the scalar exchange becomes dominant. In the scaling limit \cite{Correa:2012nk},
\begin{equation}\label{CHMSlimit}
 \vartheta\rightarrow \infty ,\qquad \lambda \,{\rm e}\,^{\vartheta}-{\rm fixed},
\end{equation}
the scalar contribution is of order one, while the gluon contribution remains small because we should simultaneously take $\lambda \rightarrow 0$. It is easy to see that any scalar line connecting the two sides of the Wilson loop is enhanced by a factor of  $\mathbf{n}_1\cdot \mathbf{n}_2=\cos\theta\sim \,{\rm e}\,^{\vartheta} $, while adding a gluon line or an internal vertex produces an uncompensated coupling $\lambda $. The scalar ladders (diagrams without internal vertices\footnote{There are also diagrams without internal vertices, where propagators connect points on one and the same line. However such diagrams cancel, since a separate straight Wilson line is supersymmetric.}) are thus the only diagrams that survive in the limit $\vartheta\rightarrow \infty $ \cite{Correa:2012nk}. Resummation of the ladder diagrams can be reduced to solving a simple Schr\"odinger equation \cite{Erickson:1999qv}.

It is convenient to introduce a rescaled coupling
\begin{equation}\label{deflambdahat}
 \hat{\lambda }=\frac{\lambda}{2}\left(\cos\theta +1\right),
\end{equation}
which is of order one in the limit (\ref{CHMSlimit}), (\ref{vartheta}), and to consider the Coulomb charge as a function of $\hat{\lambda }$ and $\lambda $. The  ladder expansion can be regarded as series in $\,{\rm e}\,^{-\vartheta}$, or equivalently in $\lambda $:
\begin{equation}
 \alpha (\lambda ,\hat{\lambda })=\sum_{n=0}^{\infty }\alpha _n(\hat{\lambda })\lambda ^n.
\end{equation}
Our goal is to compute $\alpha _1$. To make the presentation self-contained, we will also review the leading order of the ladder approximation that computes $\alpha _0$.

\section{Ladder diagrams}

\subsection{Bethe-Salpeter equation: general structure}

To calculate the $n$-th order correction to the Coulomb charge, in principle it is enough to identify all diagrams that have order $O(\lambda^n )$ in the scaling limit (\ref{CHMSlimit}). But the number of such diagrams is infinite, and a better way to organize the calculation is to first compute 2PI diagrams and then generate all other diagrams with the help of the Bethe-Salpeter equation. The number of 2PI diagrams   at each order is finite. In addition, the Bethe-Salpeter equation automatically exponentiates the result, making it easy to take the logarithm in (\ref{logW}). 

The kernel of the Bethe-Salpeter equation  $K(s',s,t',t)$ is the sum of 2PI diagrams whose end-points lie within the intervals $(s,s')$ and $(t,t')$ on the two anti-parallel sides of the Wilson loop. The sum of all diagrams, not necessarily 2PI, with end-points within the intervals $(0,S)$ and $(0,T)$, which we denote by $\Gamma (S,T)$, then satisfies the Bethe-Salpeter equation (fig.~\ref{BSEfig}):
\begin{figure}[t]
\begin{center}
 \centerline{\includegraphics[width=13cm]{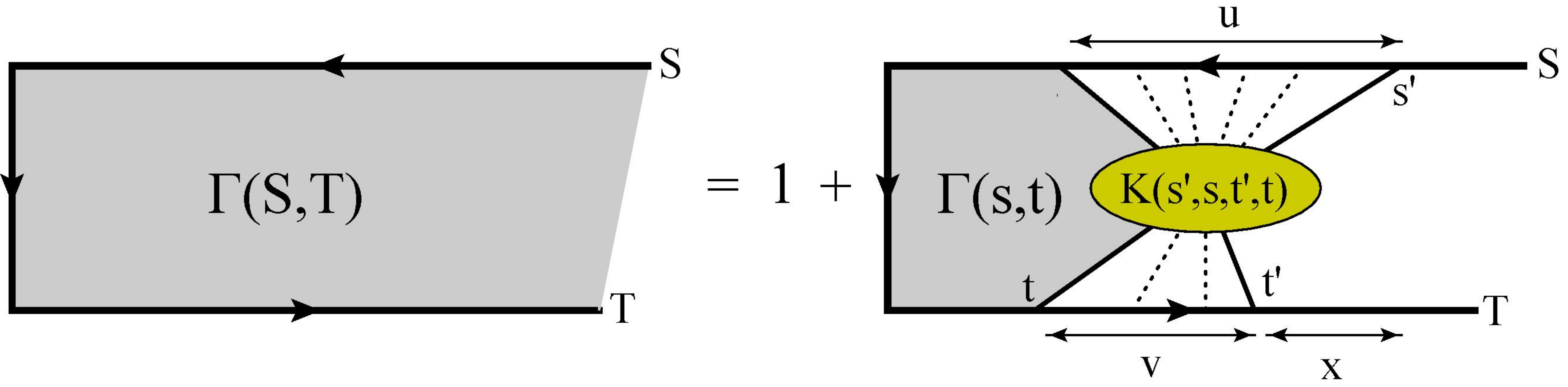}}
\caption{\label{BSEfig}\small The Bethe-Salpeter equation.}
\end{center}
\end{figure}
\begin{equation}\label{basicBS}
 \Gamma (S,T)=1+\int_{0}^{S}ds'\int_{0}^{s'}ds \int_{0}^{T}dt'\int_{0}^{t'}dt \,
 {K}(s',s;t',t )\Gamma (s ,t).
\end{equation}
The Coulomb charge is related to the large-distance behavior of $\Gamma (S,T)$: in the units where $L=1$, 
\begin{equation}\label{defa}
 \alpha =\lim_{T\rightarrow \infty }\frac{\log{\Gamma (T,T)}}{T}\,.
\end{equation}
The equation can be simplified by differentiating with respect to $S$ and $T$:
\begin{equation}
 \frac{\partial^2 \Gamma }{\partial  S\,\partial  T}
 =\int_{0}^{S}ds\int_{0}^{T}dt \,{K}(S,s ;T,t )\Gamma (s ,t ).
\end{equation}

It is convenient to introduce new variables (fig.~\ref{BSEfig}):
\begin{equation}\label{change1}
 y=S+T,\qquad x=S-T,\qquad 
 u =S-s,\qquad v =T-t.
\end{equation}
Owing to the translation invariance, the kernel does not depend on the center-of-mass coordinate $y$:
\begin{equation}
 {K}(S,s;T,t)\equiv 
 K(u,v;x ),
\end{equation}
and the Bethe-Salpeter equation becomes
\begin{equation}\label{BSeq}
 \frac{\partial ^2\Gamma }{\partial y^2}-\frac{\partial^2 \Gamma }{\partial x^2}
 =\int_{0}^{\frac{y+x}{2}}du\int_{0}^{\frac{y-x}{2}}dv \,
 K(u ,v ;x)\Gamma (y-u-v ,x-u +v).
\end{equation}
The equation should be supplemented with the boundary condition
\begin{equation}\label{bc}
 \left. \vphantom{\frac{1}{2}}\Gamma (y,x)\right|_{y=|x|}=1.
\end{equation}

According to (\ref{defa}), the Bethe-Salpeter wavefunction $\Gamma (y,x)$ grows exponentially with $y$, at least when $x=0$. Under the assumption that at small but finite $x$ the dependence on $y$ is still exponential, we arrive at the following ansatz:
\begin{equation}\label{fact}
 \Gamma (y,x)\simeq \psi (x)\,{\rm e}\,^{\frac{\alpha y}{2}}\,,
\end{equation}
where $\alpha $ is the Coulomb charge that we want to calculate. This ansatz can be valid only at $|x|\ll y$, as is does not carry any information on the boundary condition (\ref{bc}). At the leading ladder approximation it is possible to make a more rigorous argument and derive the behavior (\ref{fact}) from the first principles \cite{Erickson:2000af}. We will not go into such a detail here, and will simply assume that (\ref{fact}) holds for sufficiently large $y$. This assumption will be sufficient for finding both $\alpha $ and $\psi (x)$.

Substituting  the ansatz (\ref{fact})  into the Bethe-Salpeter equation (\ref{BSeq}), and sending $y$ to infinity while keeping $x$ finite, we get an integro-differential equation for $\psi (x)$:
\begin{equation}\label{eigenvalue}
 \frac{d^2\psi }{d x^2}+
 \int_{0}^{\infty }du\int_{0}^{\infty }dv \,
 \,{\rm e}\,^{-\frac{\alpha }{2}\left(u+v \right)}
 K(u,v ;x)\psi (x-u+v )=\frac{\alpha ^2}{4}\,\psi .
\end{equation}
The Coulomb charge $\alpha $ is the smallest eigenvalue of this linear problem. 

\subsection{Summing up ladders}

The Bethe-Salpeter kernel can be computed in perturbation theory:
\begin{equation}
 K(u,v;x)=\sum_{n=0}^{\infty }\lambda ^nK_n(u,v;x),
\end{equation}
where each order contains a finite number of 2PI diagrams. Each $K_n$ is a function of $\hat{\lambda }$.

The only 2PI diagram of order $O(\lambda ^0)$ is the scalar exchange:
\begin{equation}\label{Ktree}
 K_0(u,v ;x)=\frac{\hat{\lambda }\delta (u )\delta (v)}{4\pi ^2\left(x^2+1\right)}\,.
\end{equation}
With this approximation,
the eigenvalue problem (\ref{eigenvalue}) reduces to an ordinary Schr\"odinger equation:
\begin{equation}\label{Sch}
  \frac{d ^2\psi_0 }{d x^2}+\frac{\hat{\lambda }}{4\pi ^2\left(x^2+1\right)}\,\psi_0 =\frac{\alpha ^2_0}{4}\,\psi_0 ,
\end{equation}
which has been analyzed in great detail in \cite{Erickson:1999qv,Erickson:2000af,Klebanov:2006jj,Correa:2012nk}. In particular, the eigenvalue of the Schr\"odinger operator can be systematically analyzed at large and at small $\hat{\lambda }$. For completeness, we review the  results below.

The weak-coupling expansion of $\alpha  _0$ is not exactly a power series: the Taylor coefficients contain logarithms. Such a non-analytic behavior arises because of the IR divergences of perturbation theory for the static potential \cite{Appelquist:1977es}. The logarithms first show up at two loops \cite{Erickson:1999qv}, and can be resummed using a version of renormalization group equation \cite{Pineda:2007kz}. Both logarithmic and finite terms can be calculated from the Schr\"odinger equation. To the three-loop accuracy \cite{Correa:2012nk}:
\begin{equation}\label{weakalpha0}
 \alpha _0(\hat{\lambda })
 =\frac{\hat{\lambda }}{4\pi }
 +\frac{\hat{\lambda }^2}{8\pi ^3}\,\ln\frac{\hat{\lambda }\,{\rm e}\,^{\gamma-1 }}{2\pi } 
 +\frac{\hat{\lambda }^3}{32\pi ^5}\left(
 \ln\frac{\hat{\lambda }\,{\rm e}\,^{\gamma }}{2\pi }
\,\,\ln\frac{\hat{\lambda }\,{\rm e}\,^{\gamma +1}}{2\pi }
-\frac{7}{2}-\frac{\pi ^2}{12}
 \right)+\ldots ,
\end{equation}
where $\gamma $ is the Euler constant.

At large $\hat{\lambda }$, the eigenvalue of the Schr\"odinger problem (\ref{Sch}) is given by the classical energy at the minimum of the potential \cite{Erickson:1999qv}:
\begin{equation}\label{strongalpha0}
 \alpha _0=\frac{\sqrt{\hat{\lambda }}}{\pi }-1+\ldots .
\end{equation}
The second term is the energy of the zero-point fluctuations. The strong-coupling asymptotics of  the ladder diagrams perfectly agrees with the classical calculation in string theory, as soon as the correct range of parameters is identified  \cite{Correa:2012nk}. The angle $\theta $ defines the boundary conditions for the string worldsheet on $S^5$ in the $AdS_5\times S^5$ geometry. Once the angle is analytically continued to complex values and
$\vartheta$ is taken to infinity, the area of the minimal surface in $AdS_5\times S^5$ grows exponentially, as $\,{\rm e}\,^{\vartheta/2}$, reproducing the ladder result (\ref{strongalpha0}).
It would be very interesting to compare the second term with the one-loop correction due to string fluctuations, which can be calculated along the lines of \cite{Drukker:2011za}\footnote{We thank Nadav Drukker for the discussion of this point.}.

The wavefunction in the strong-coupling limit is highly peaked near zero:
\begin{equation}\label{spilarge}
 \psi _0(x)\simeq \left(\frac{\hat{\lambda} }{4\pi ^4}\right)^{\frac{1}{8}}\,{\rm e}\,^{-\frac{\sqrt{\hat{\lambda} }\,x^2}{4\pi }}\qquad \left(\lambda \rightarrow \infty \right).
\end{equation}
The ladder diagrams are thus getting densely packed at large $\hat{\lambda }$ \cite{Shuryak:fk}. Indeed, the average $\left\langle x^2\right\rangle\sim {\hat{\lambda }}^{-1/2}$ determines the mean square distance between consecutive rungs of the ladder. At strong coupling this distance becomes parametrically small. 

\begin{figure}[t]
\begin{center}
\subfigure[]{
   \includegraphics[width=6cm] {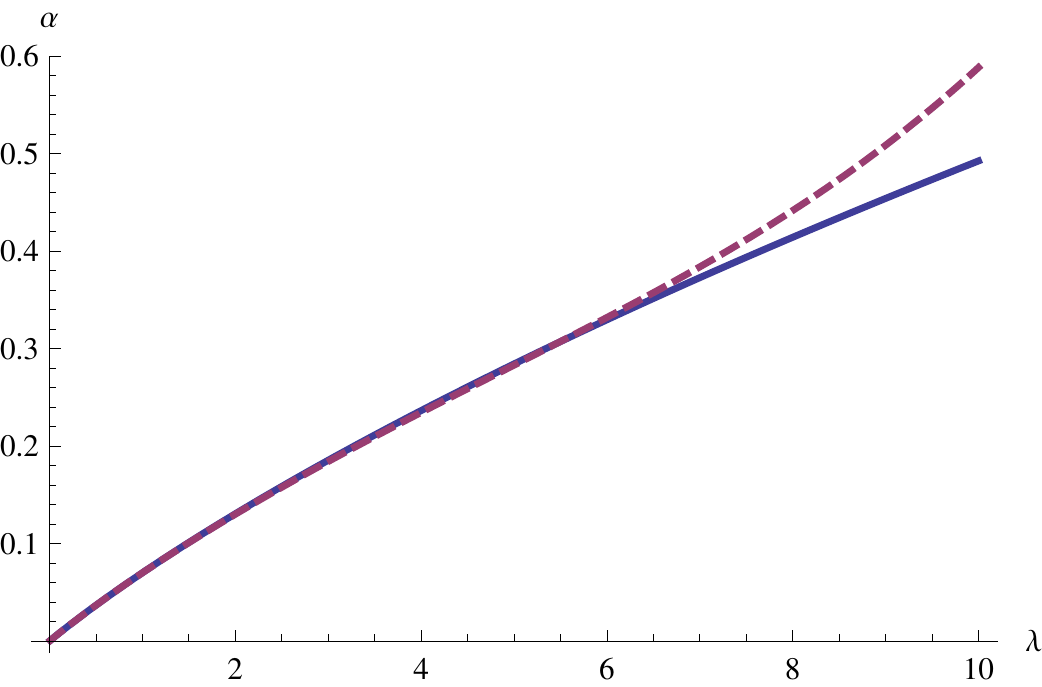}
   \label{fig:subfig1}
 }
 \subfigure[]{
   \includegraphics[width=6cm] {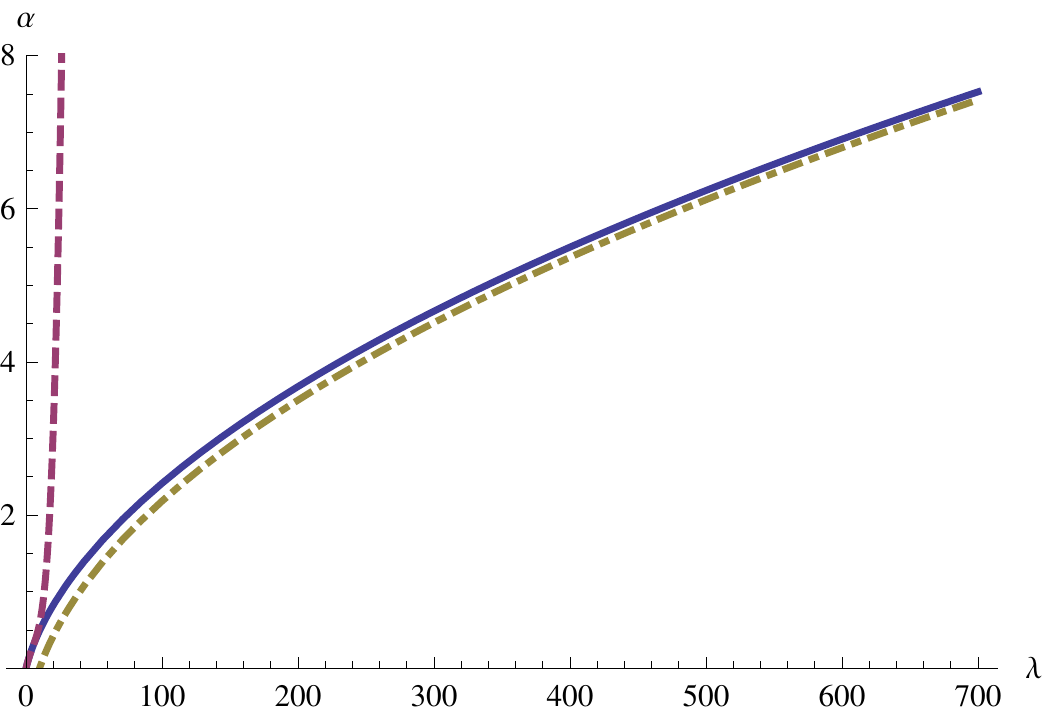}
   \label{fig:subfig2}
 }
\caption{\label{graphs}\small Comparison of the exact eigenvalue of the Schr\"odinger equation (\ref{Sch}) (solid blue line) with the three-loop weak-coupling result (\ref{weakalpha0}) (dashed magenta line) and one-loop strong-coupling approximation (\ref{strongalpha0}) (dot-dashed yellow line).}
\end{center}
\end{figure}

In fig.~\ref{graphs} the numerical solution for $\alpha _0$ in the full range of $\hat{\lambda }$ is compared to the weak and strong coupling approximations above.

\section{Beyond ladders}

\begin{figure}[t]
\begin{center}
\subfigure[]{
 {\includegraphics[width=2.2cm]{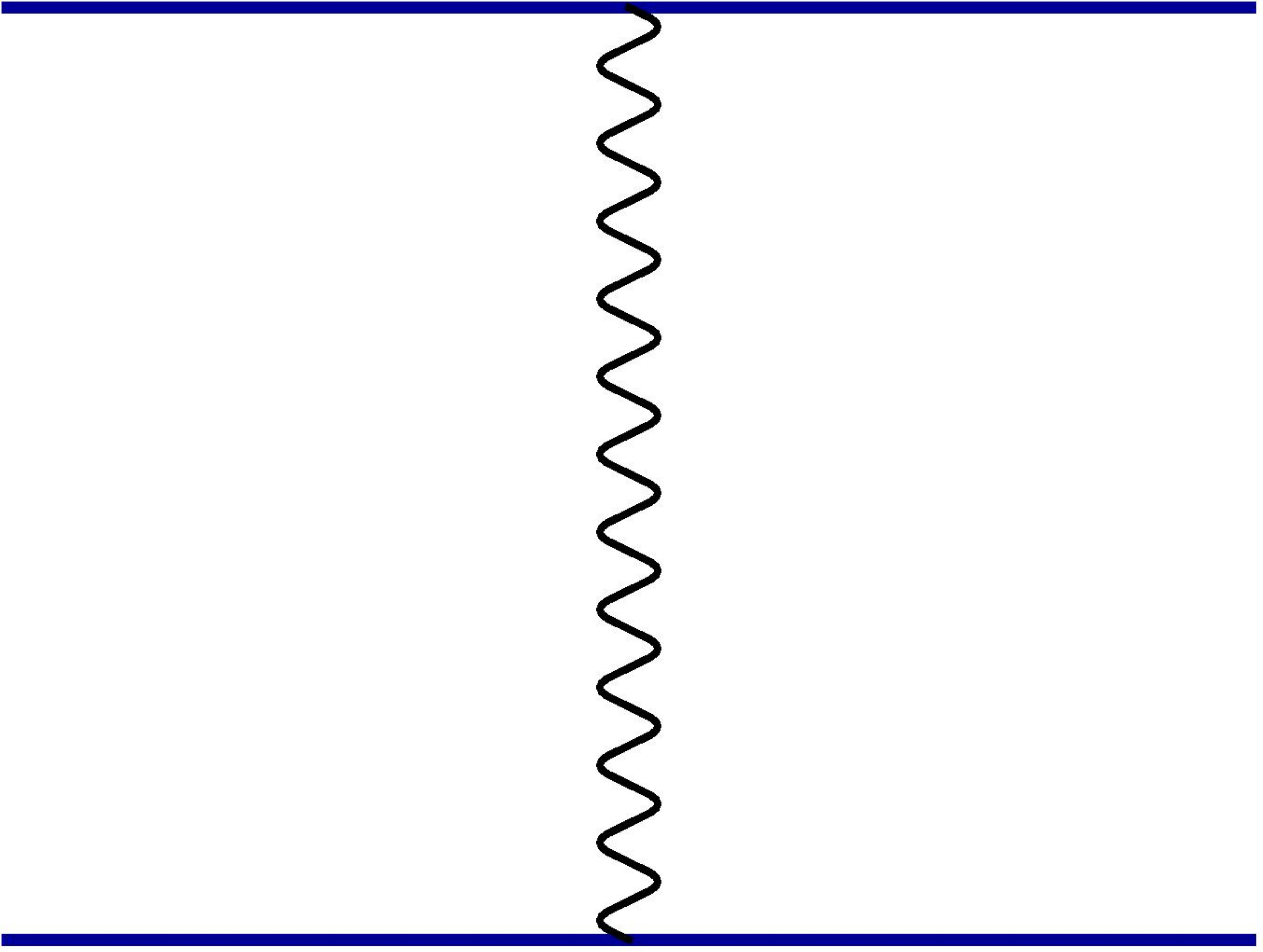}}
 }
 \subfigure[]{
 {\includegraphics[width=2.2cm]{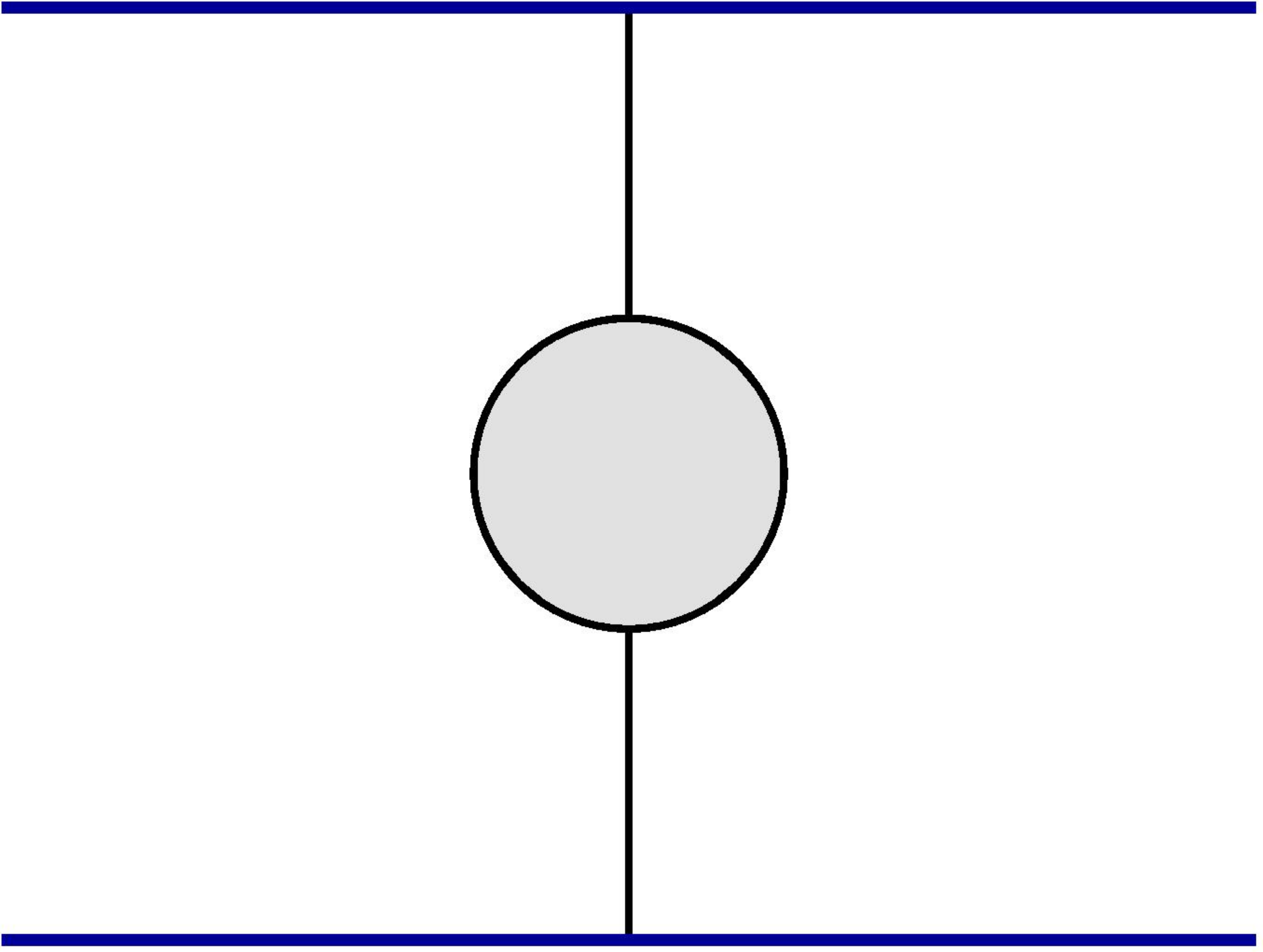}}
 }
 \subfigure[]{
 {\includegraphics[width=2.2cm]{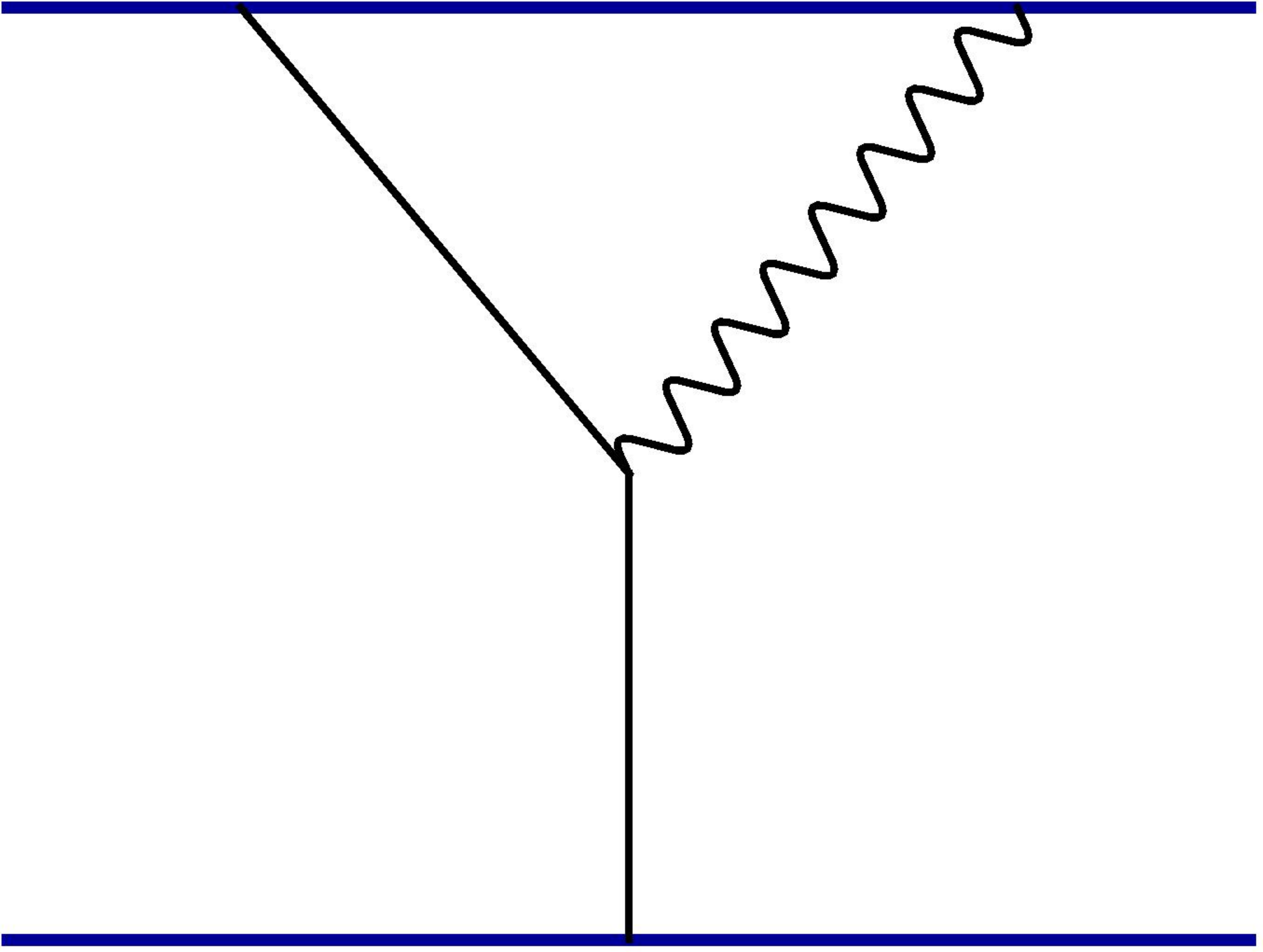}}
 }
 \subfigure[]{
 {\includegraphics[width=2.2cm]{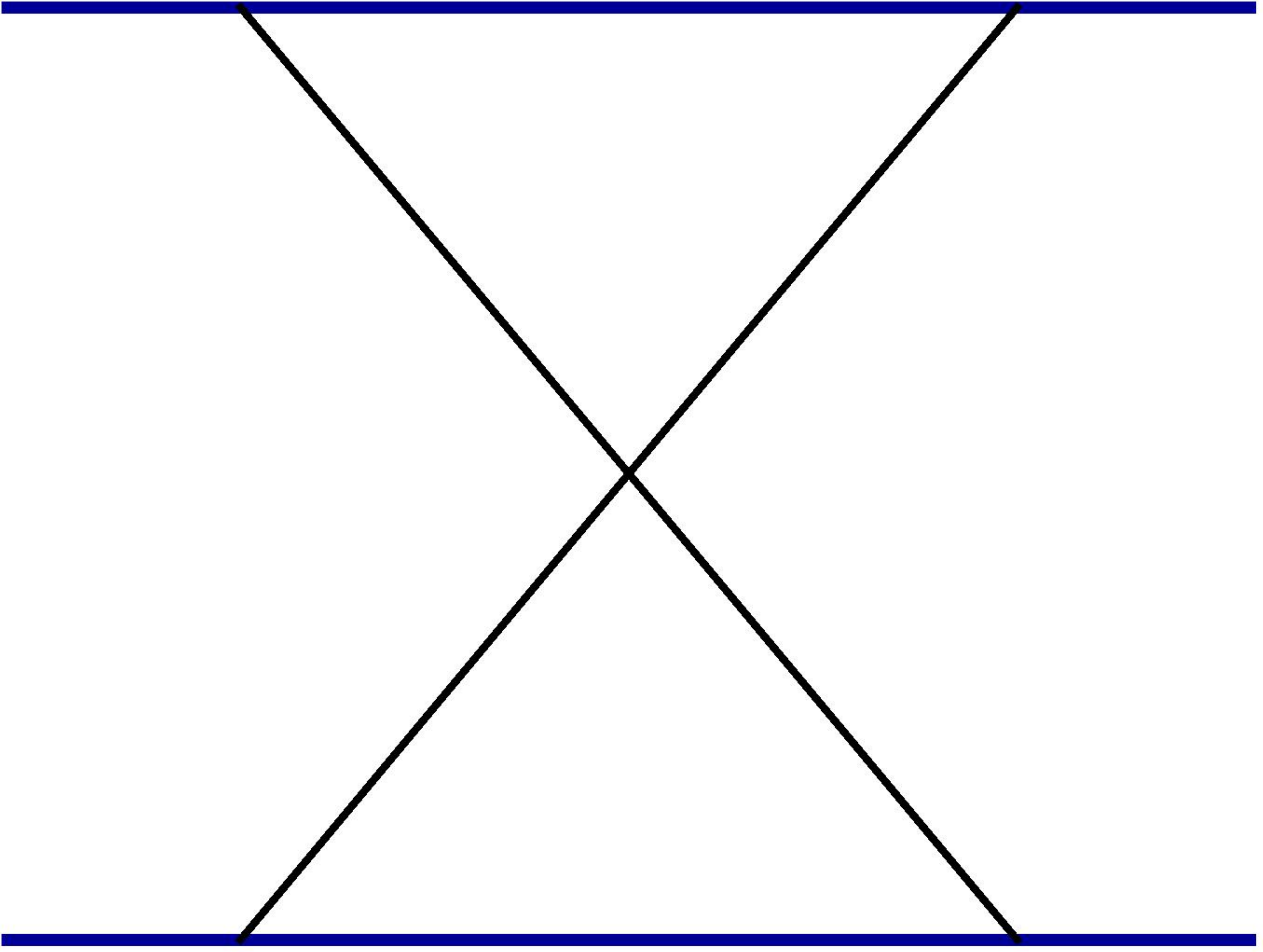}}
 }
 \subfigure[]{
 {\includegraphics[width=2.2cm]{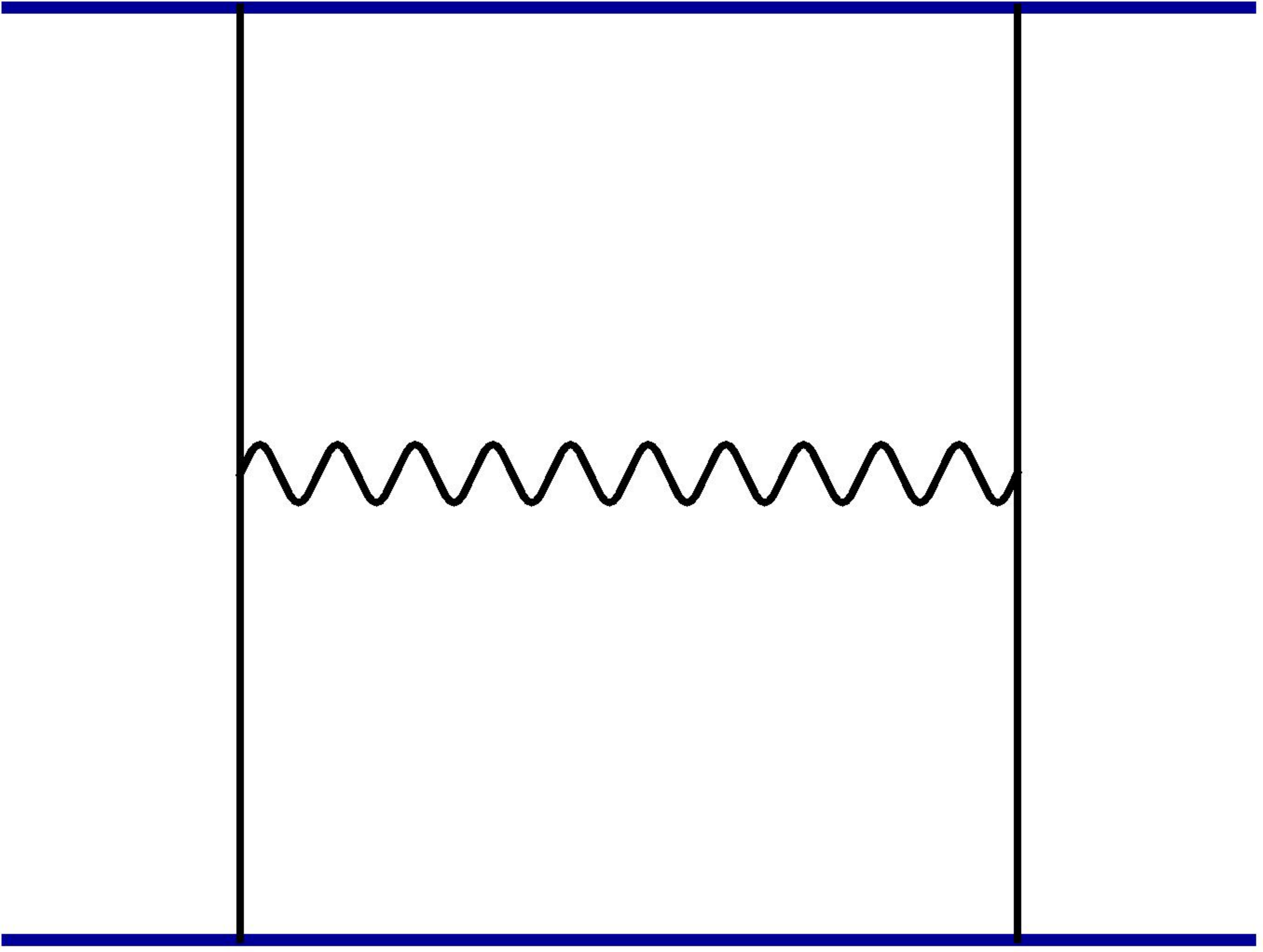}}
 }
\caption{\label{fig1-allD}\small Bethe-Salpeter kernel at the next-to-leading order.}
\end{center}
\end{figure}

To compute the NLO contribution to the Coulomb charge, we first need to calculate the $O(\lambda )$ correction to the kernel -- the sum of 2PI diagrams proportional to $\lambda $, $\lambda ^2\cos\theta $ and $\lambda ^3\cos^2\theta $ (there are no 2PI diagrams of higher order in $\hat{\lambda }$). The relevant diagrams are shown in  fig.~\ref{fig1-allD}. 
The resulting corrections to the Schr\"odinger equation (\ref{Sch}) are non-local, 
which definitely complicates the problem. On the other hand, we do not need to solve the equation exactly,
but can apply ordinary quantum-mechanical perturbation theory. As we shall see, the problem really simplifies when $\hat{\lambda }$ is very large (at strong coupling) and when $\hat{\lambda }$ is very small (at weak coupling).

Although the rectangular Wilson loop is not a supersymmetric observable, the  diagrams in fig.~\ref{fig1-allD}, when combined together, feature spectacular cancellations. In the Feynman gauge the gluon and the scalar propagators are equal. The  diagram~(a) then renormalizes the effective coupling, and it has been already taken into account by including the $1$ in the definition of $\hat{\lambda }$,  eq.~(\ref{deflambdahat}). The sum of the remaining three diagrams can be considerably simplified.

\begin{figure}[t]
\begin{center}
\subfigure[]{
 {\includegraphics[height=3.5cm]{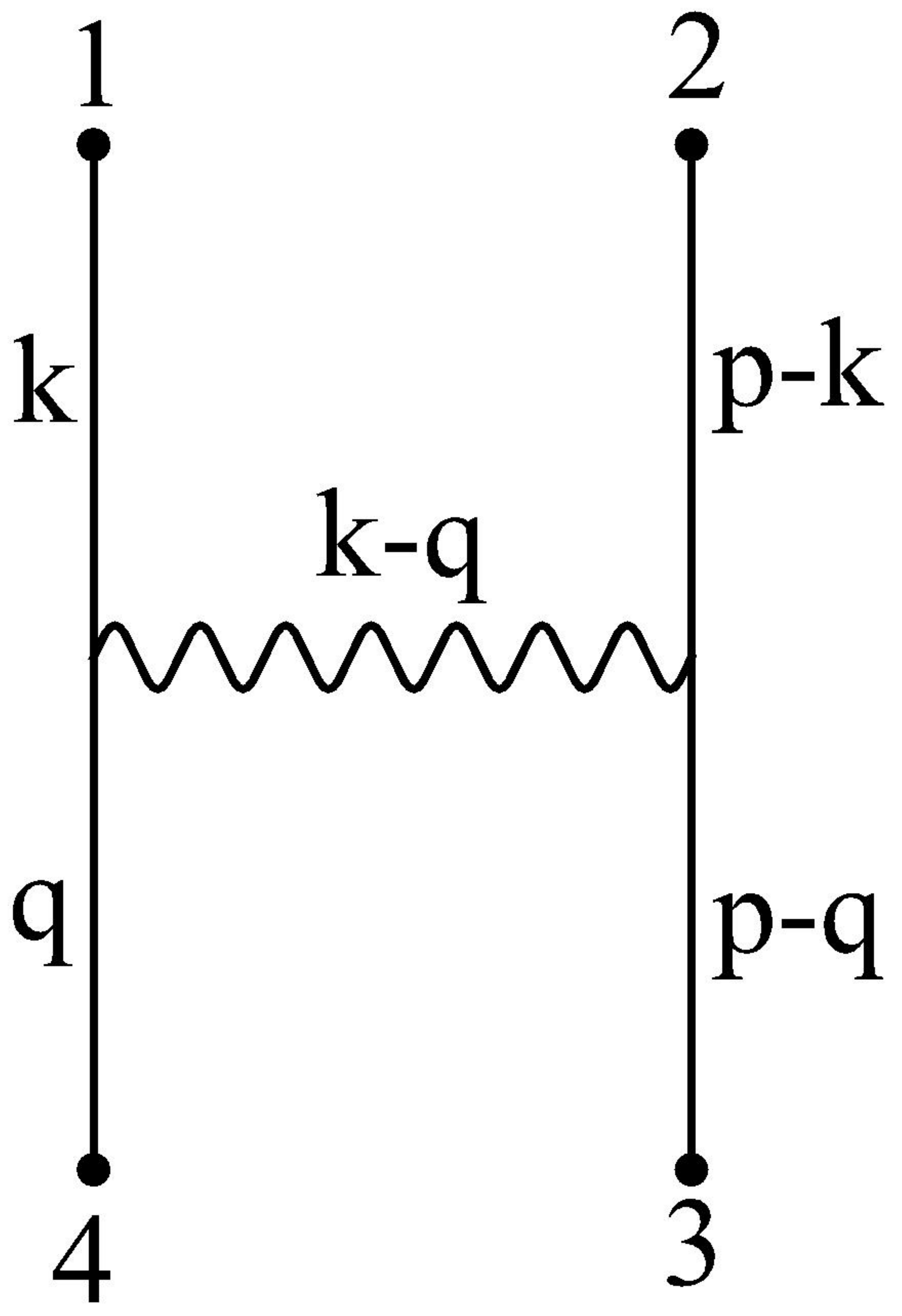}}
 }
\qquad  
 \subfigure[]{
 {\includegraphics[height=3.5cm]{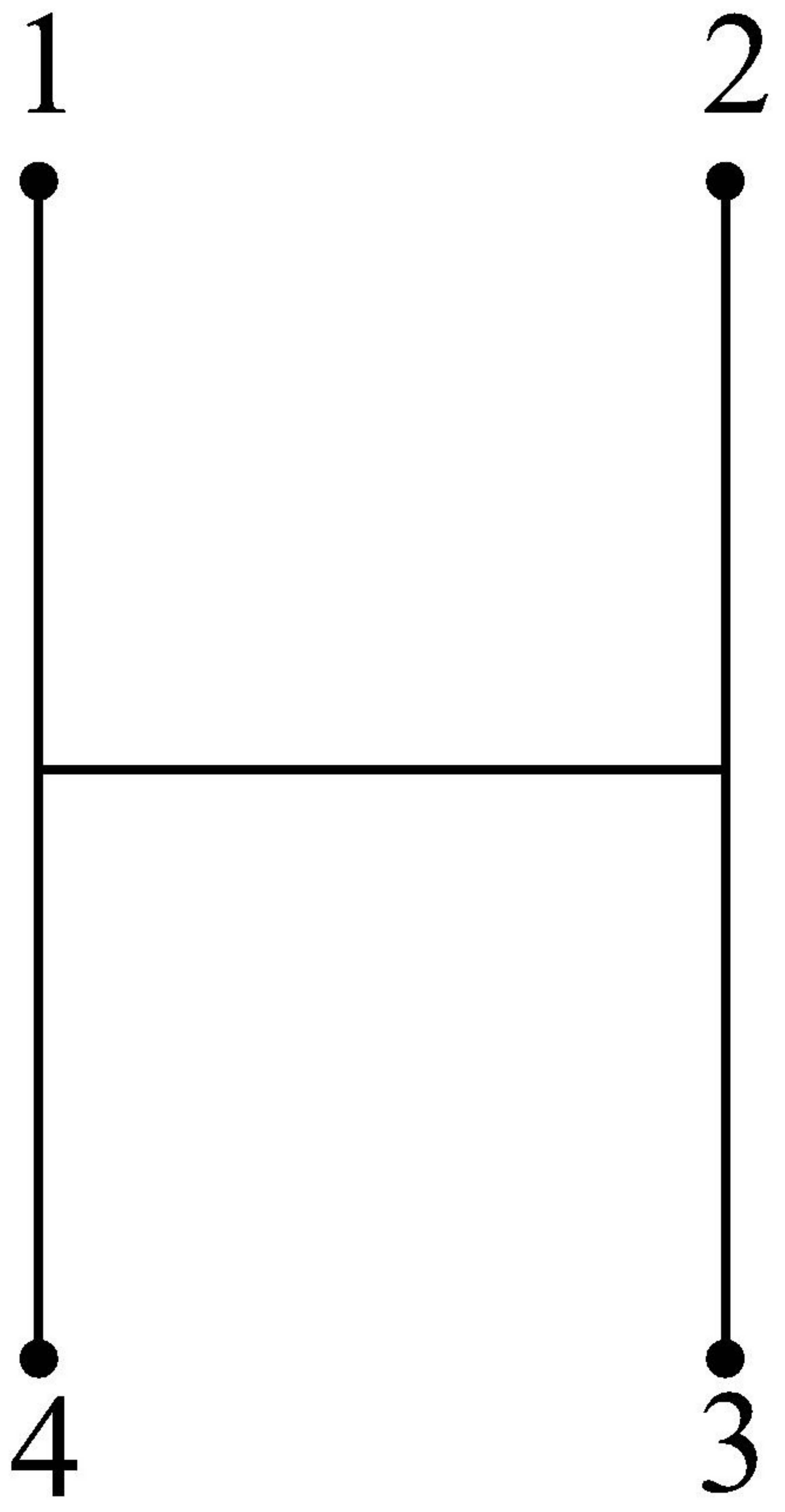}}
 }
\caption{\label{HD}\small The H-diagram.}
\end{center}
\end{figure}

We start with the H-graph, fig.~\ref{fig1-allD}(e). The numerator in the corresponding momentum integral (the momentum assignment is shown in fig.~\ref{HD}a), can be written as 
\begin{equation}
- (k+q)(2p-k-q)=-2p^2+k^2+q^2+(p-k)^2+(p-q)^2-(k-q)^2.
\end{equation}
This is the standard Passarino-Veltman reduction. All the terms but one contain a vanishing propagator. Those terms combine into
\begin{equation}\label{Passarino-Veltman}
\pictext{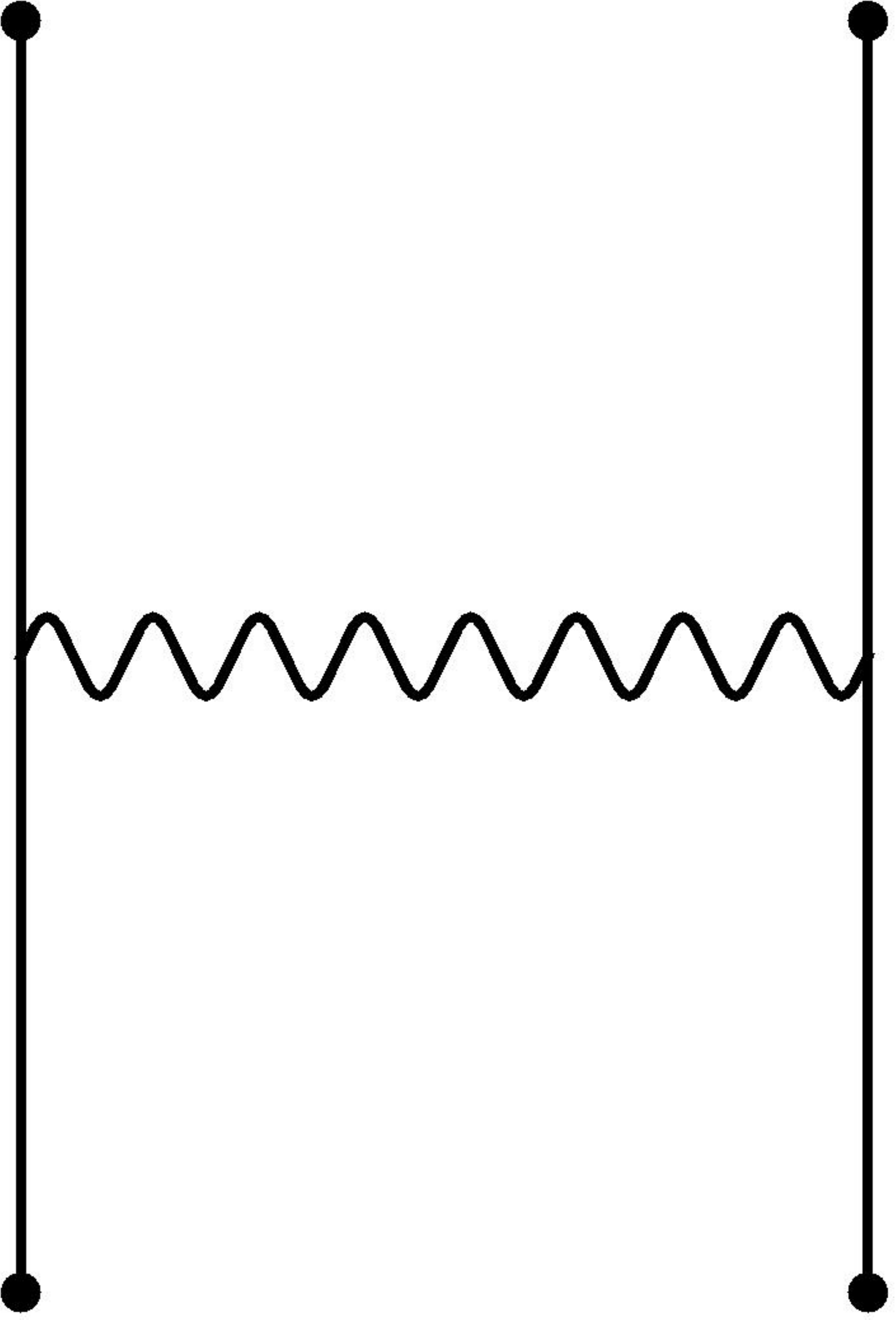}=
 2\left(\partial _1+\partial _2\right)^2\pictext{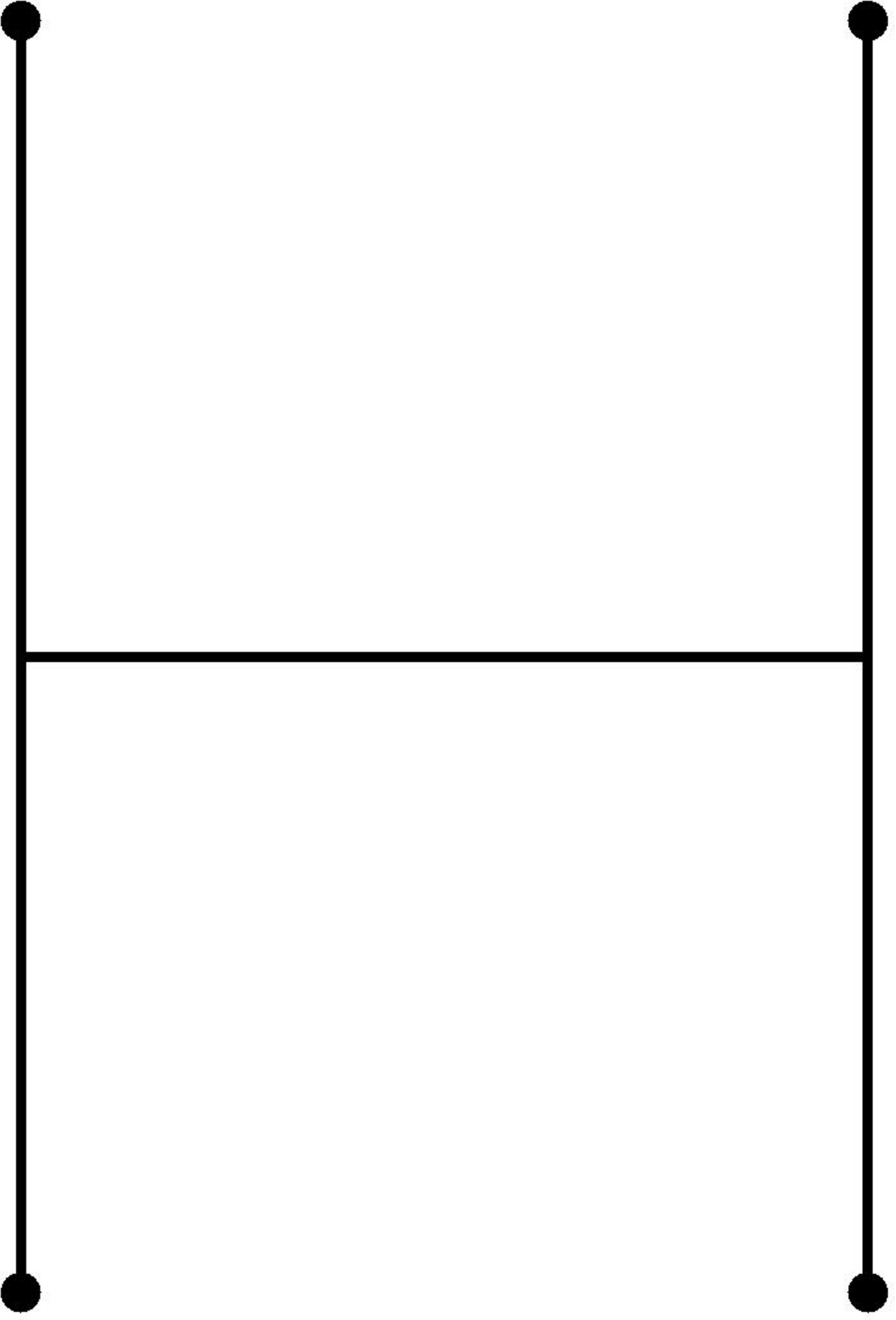}
 +\pictext{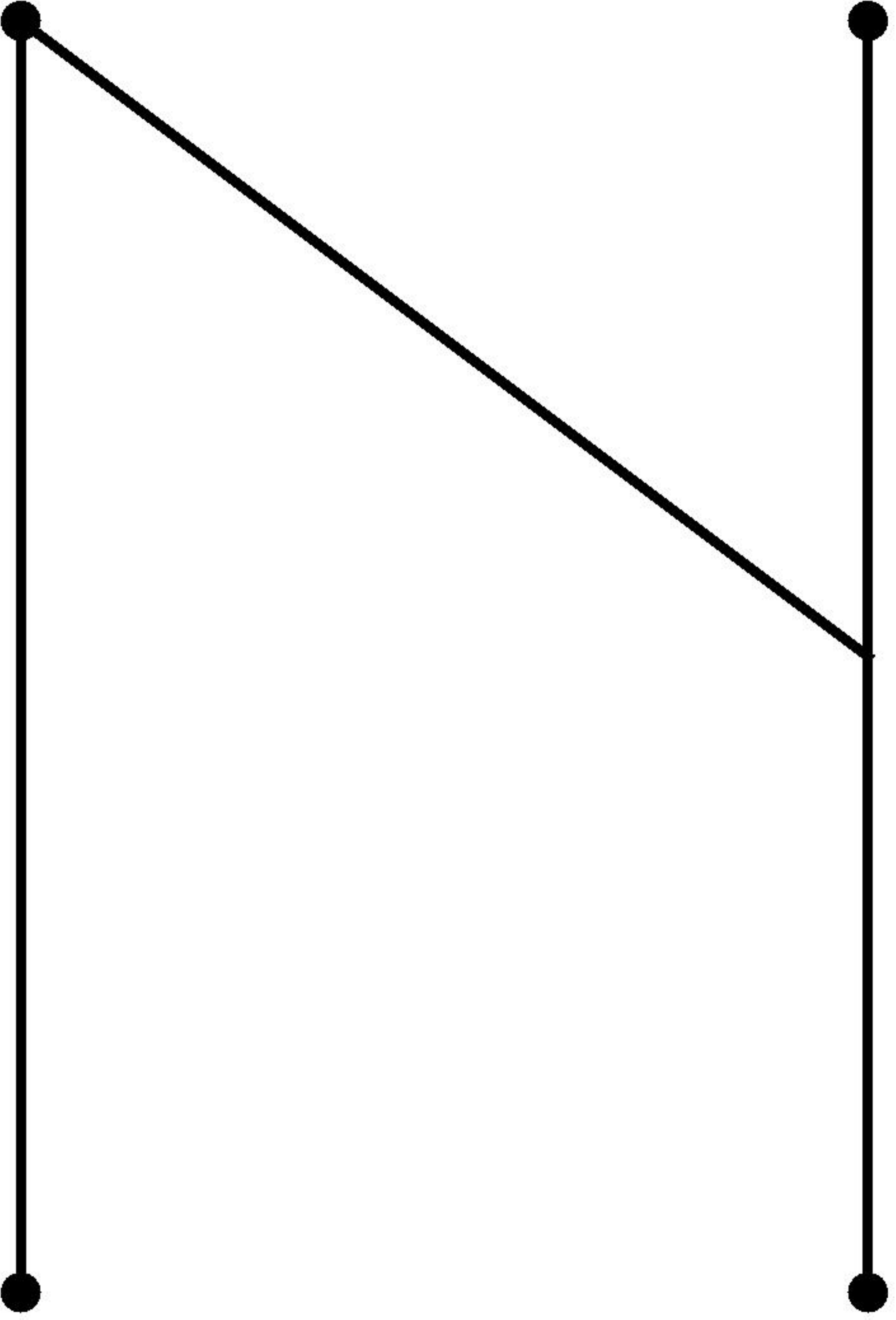}
  +\pictext{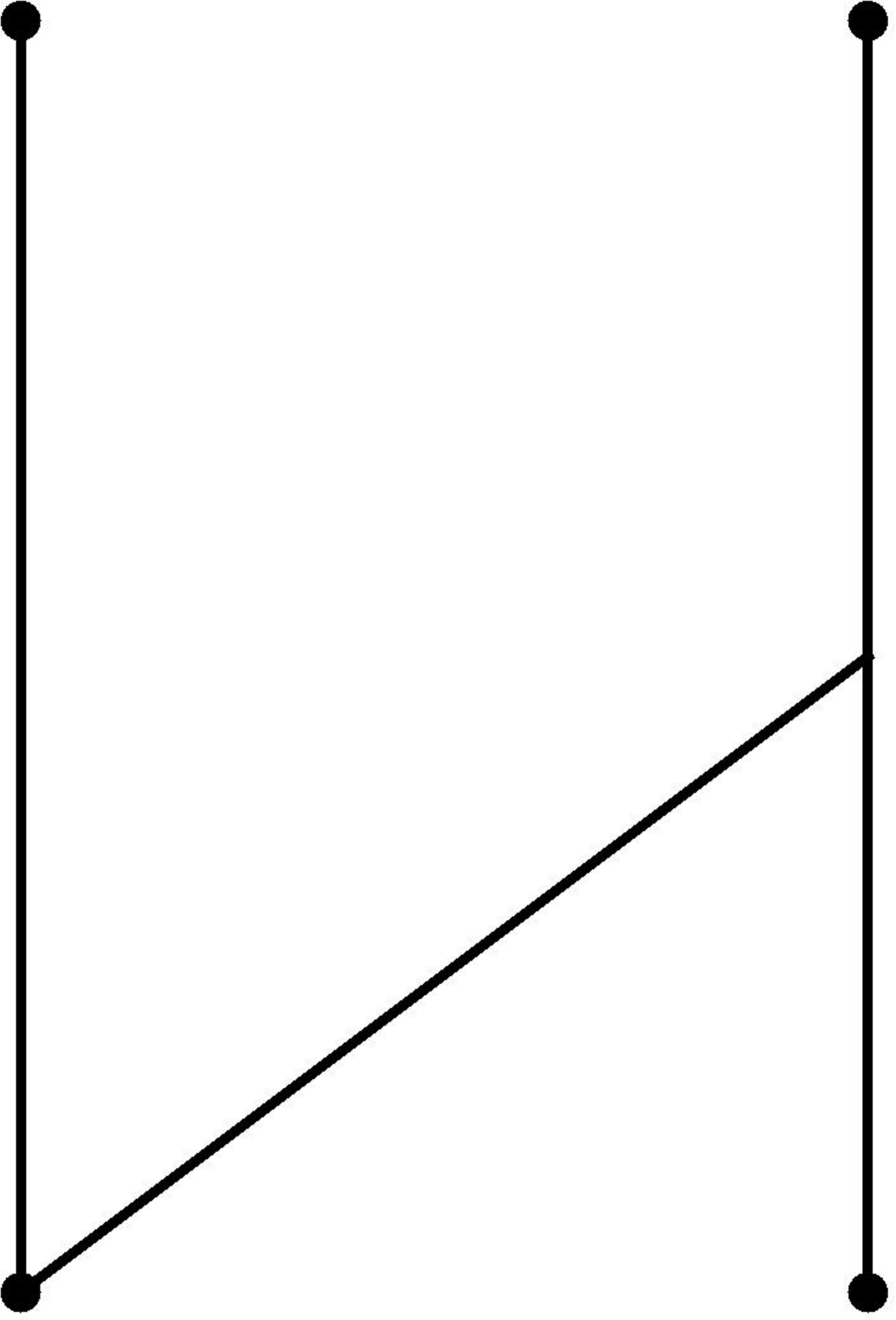}
   +\pictext{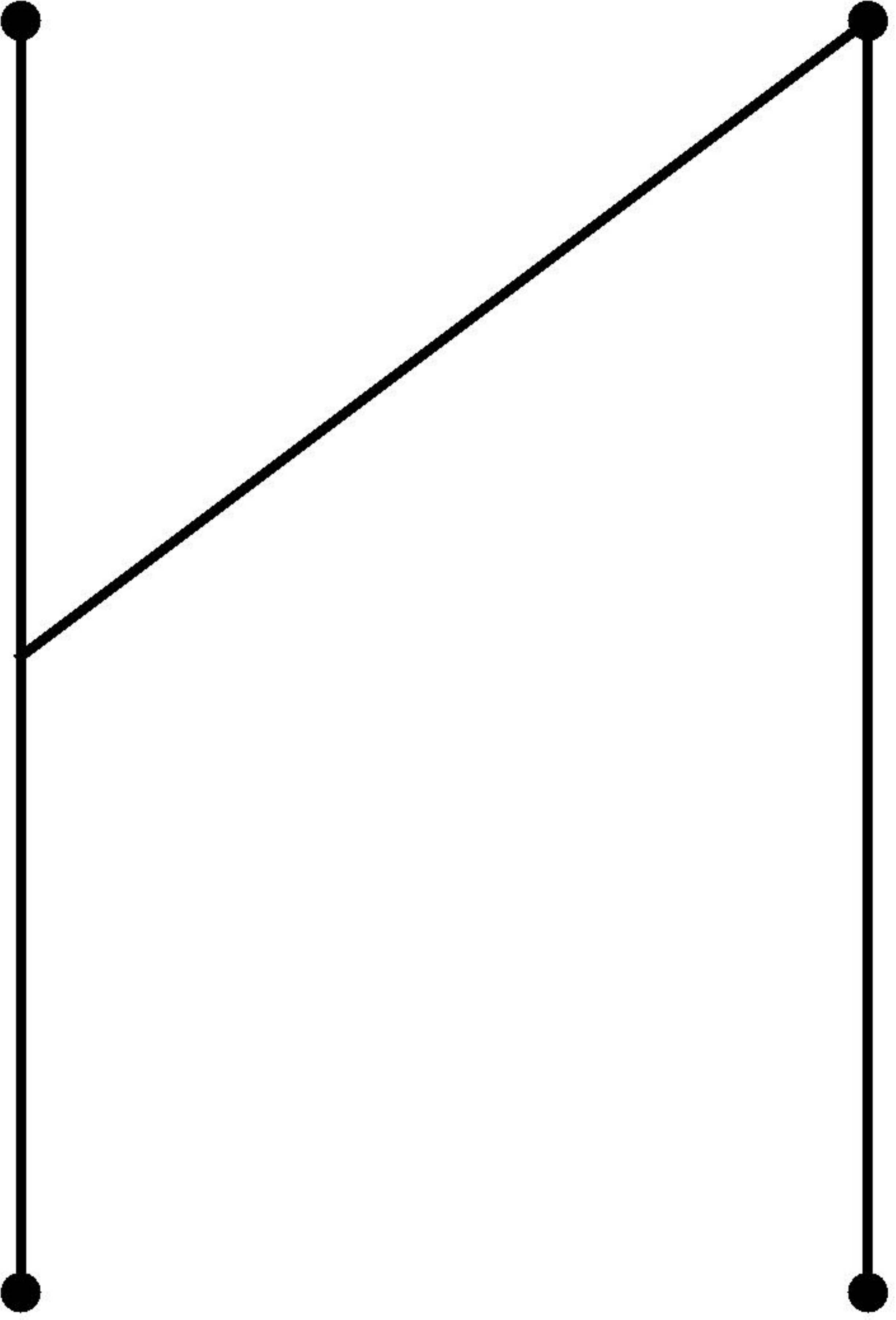}
    +\pictext{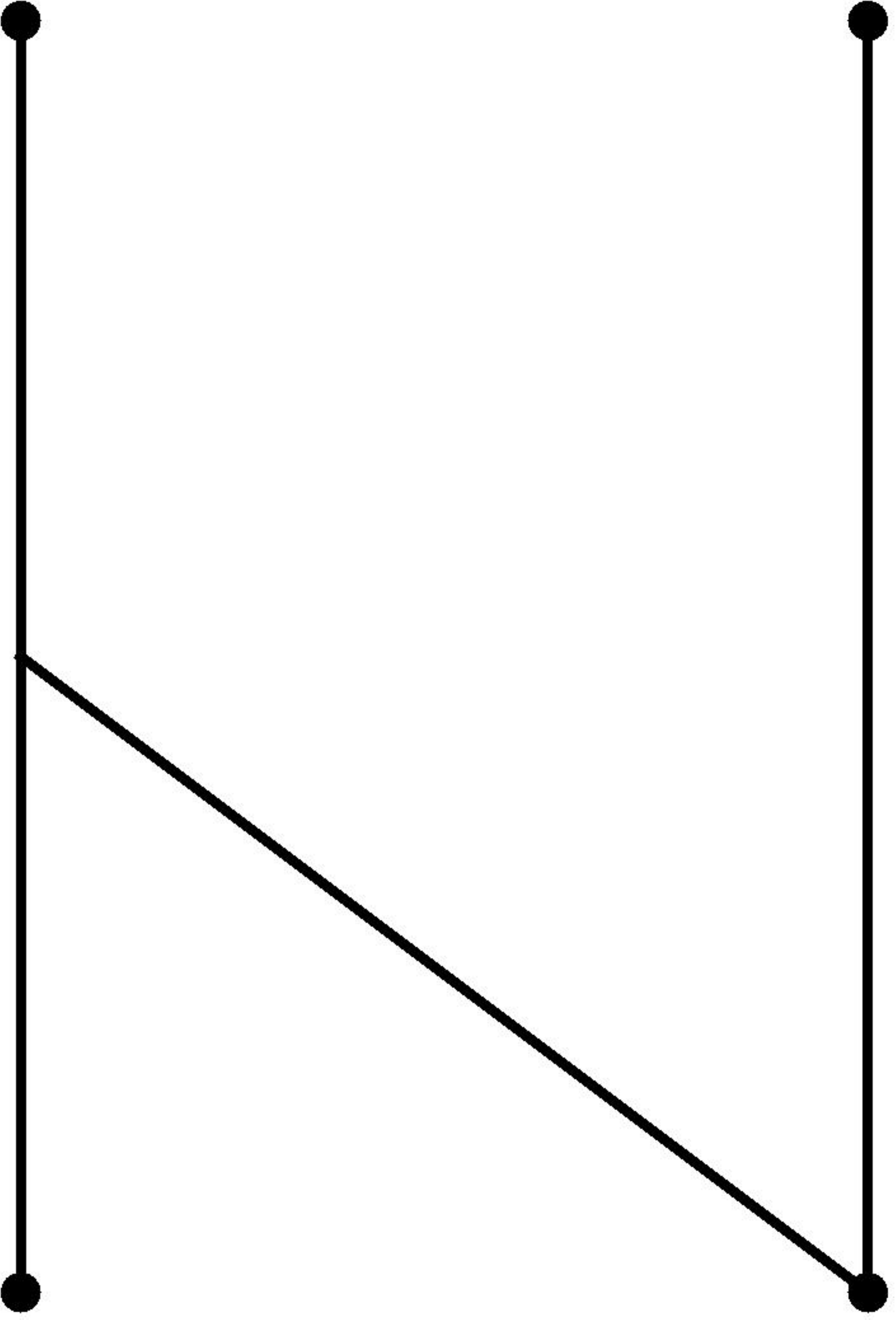}
    - \pictext{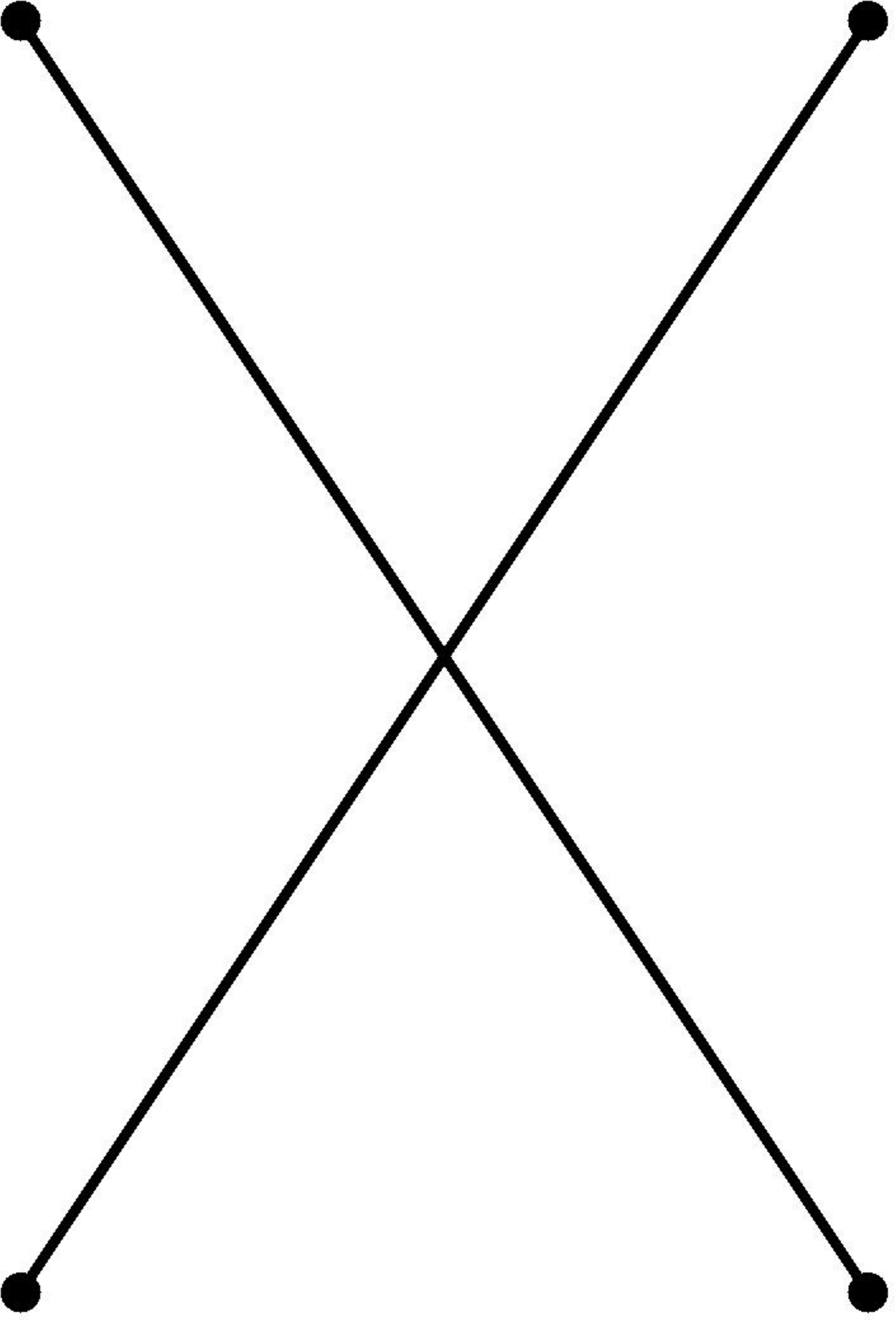}
\end{equation}
The last term exactly cancels with the X-diagram in fig.~\ref{fig1-allD}d. 

The Y-graph in  fig.~\ref{fig1-allD}(c)  can be reduced to the scalar diagram with the derivatives along the contour, acting on the external lines:
\begin{equation}
 \pictext{Ygraph.pdf}+\pictext{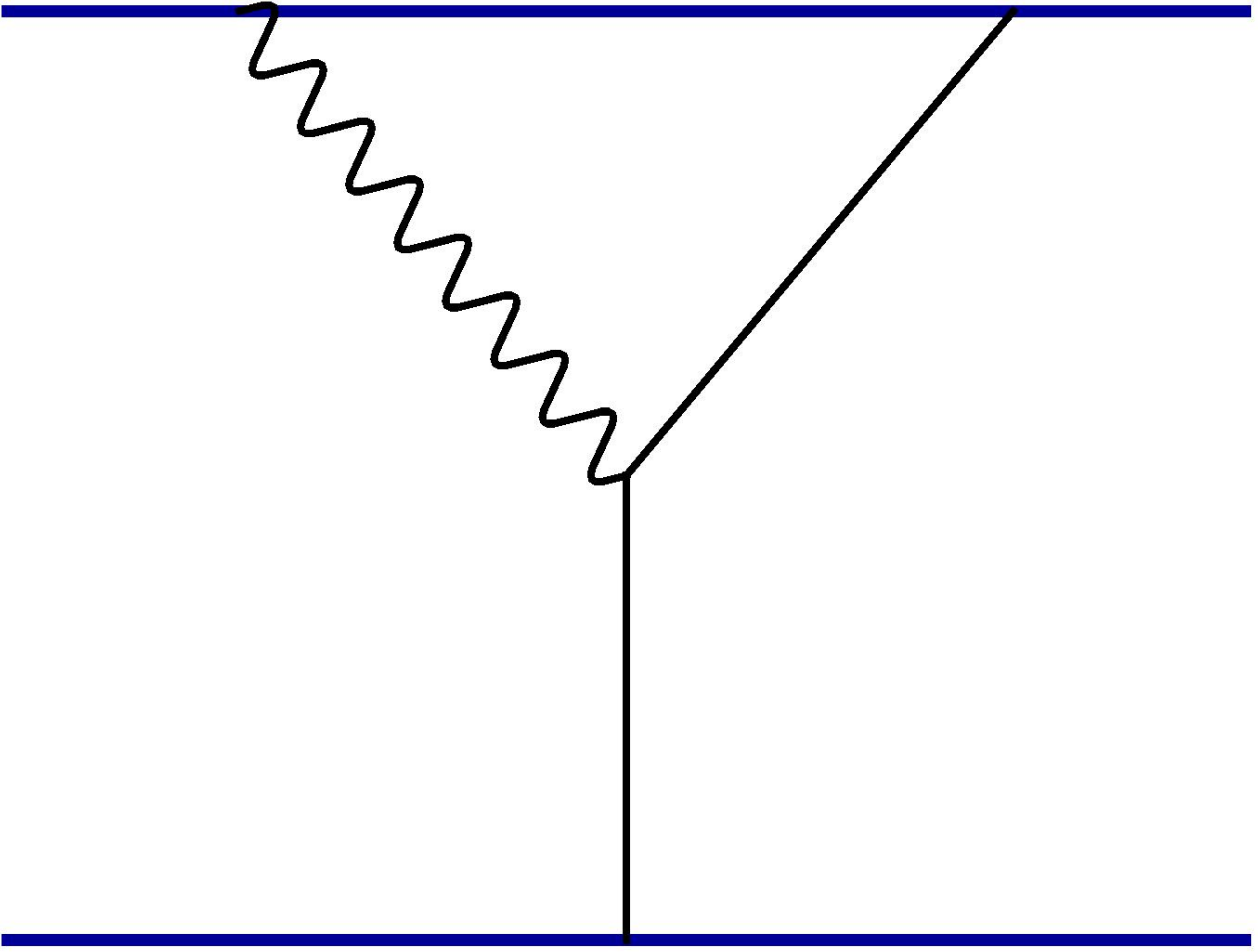}
 =\pictext{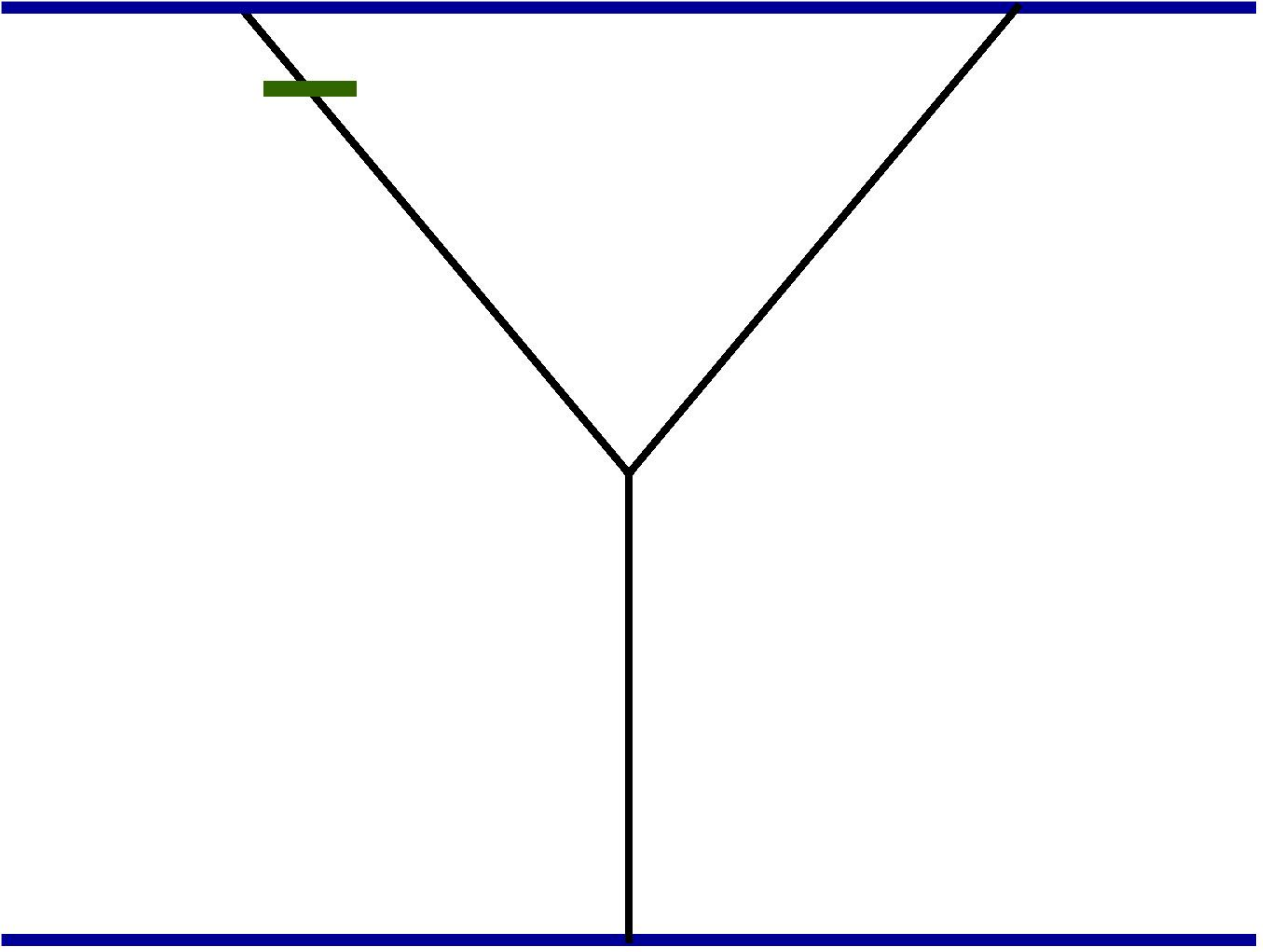}-\pictext{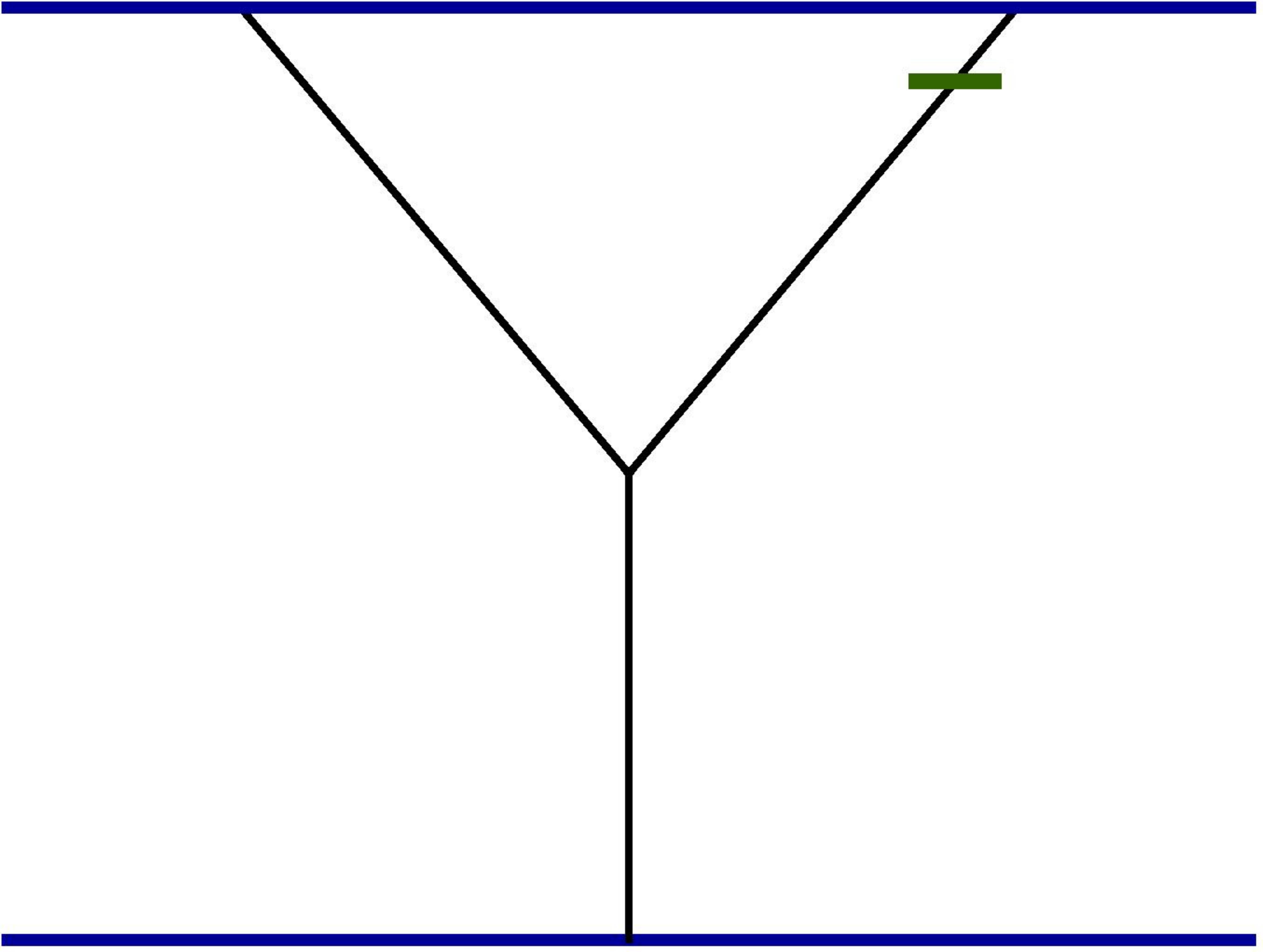}
\end{equation}
after which the external leg of the diagram can be integrated by parts. The fact that the Y-diagrams gives a total derivative has been observed many times in similar calculations \cite{Erickson:2000af,Makeenko:2006ds}. Upon integration by parts  only the boundary terms survive: 
\begin{equation}\label{Y's}
 \pictext{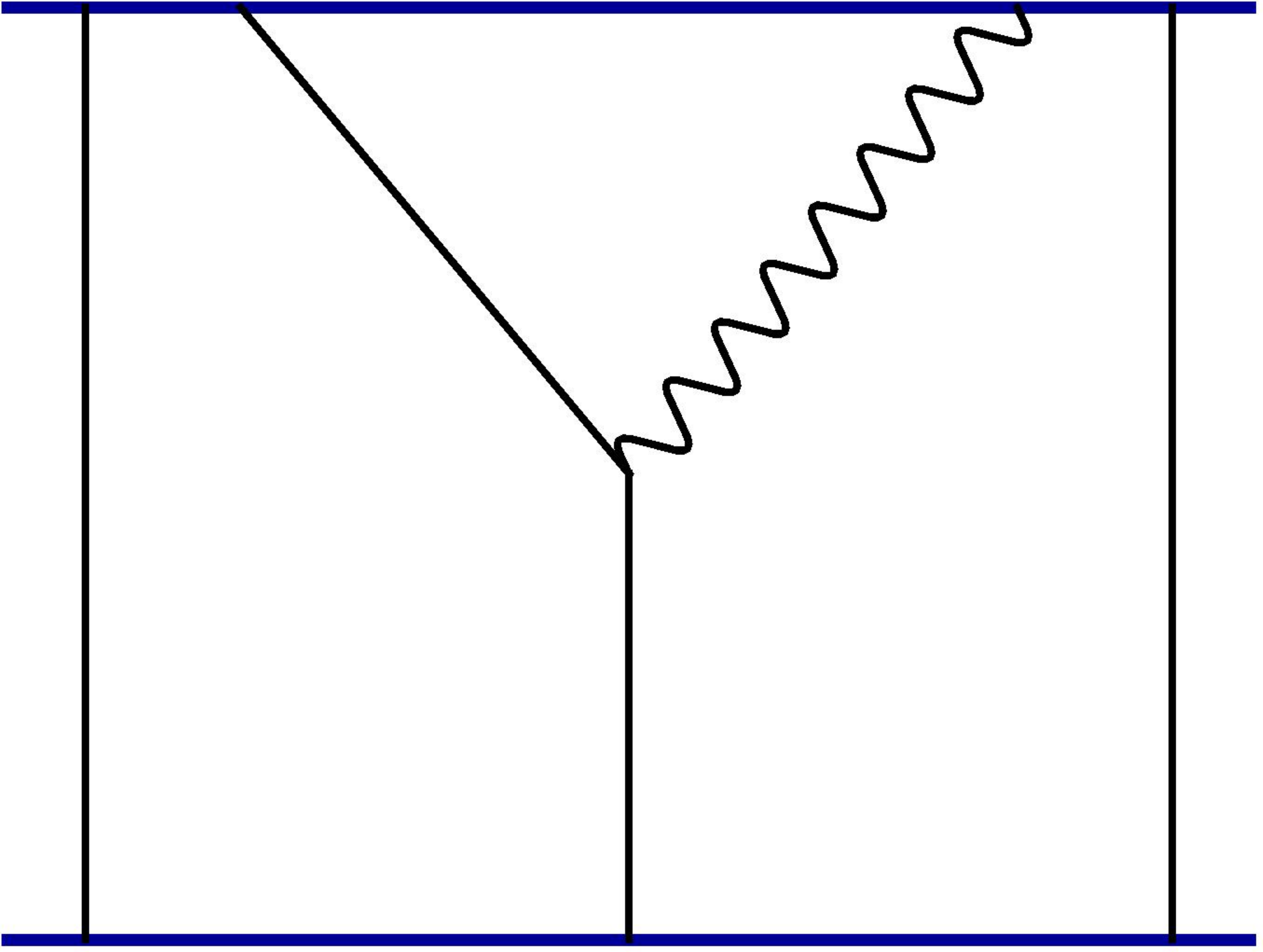}+\pictext{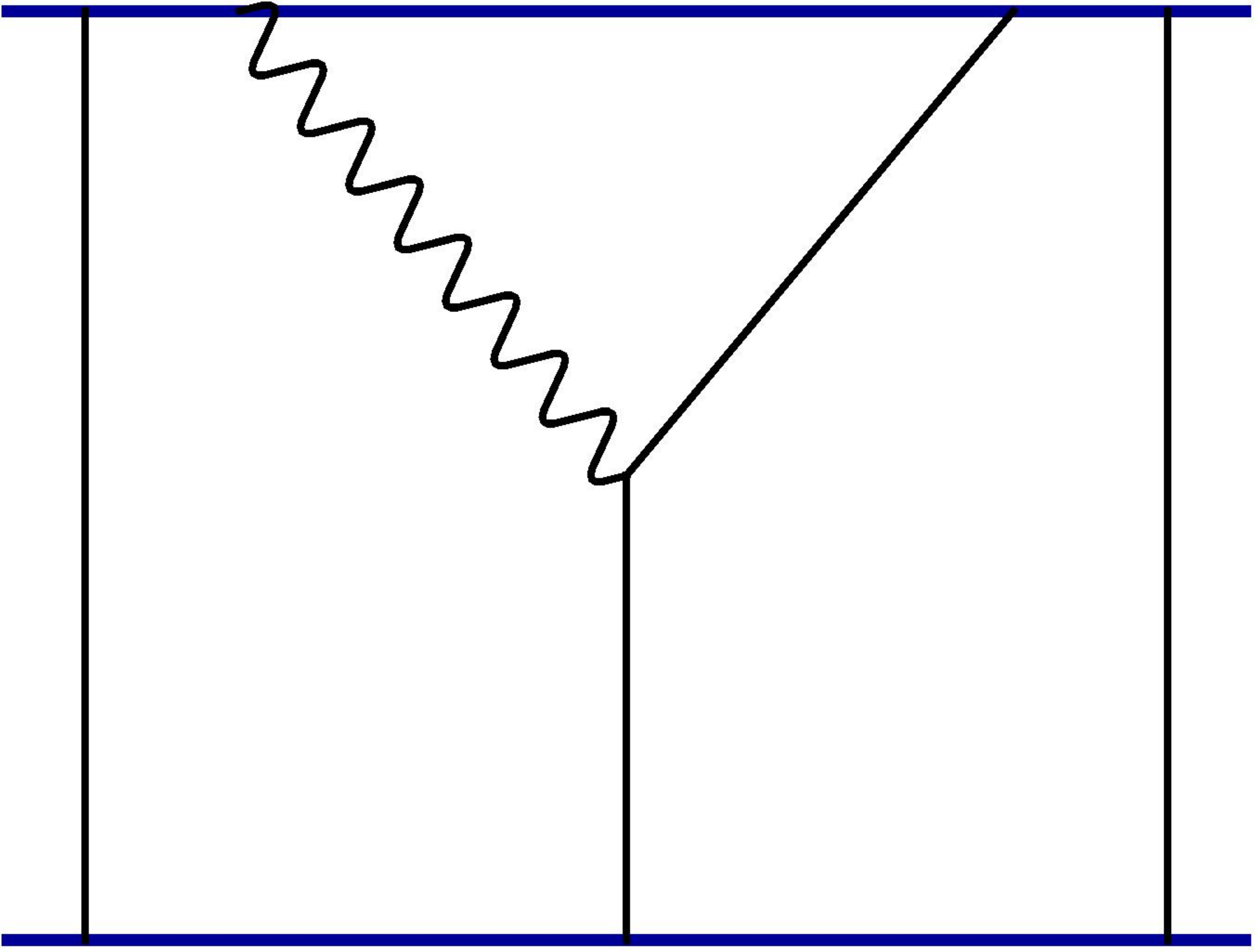}
 =2\pictext{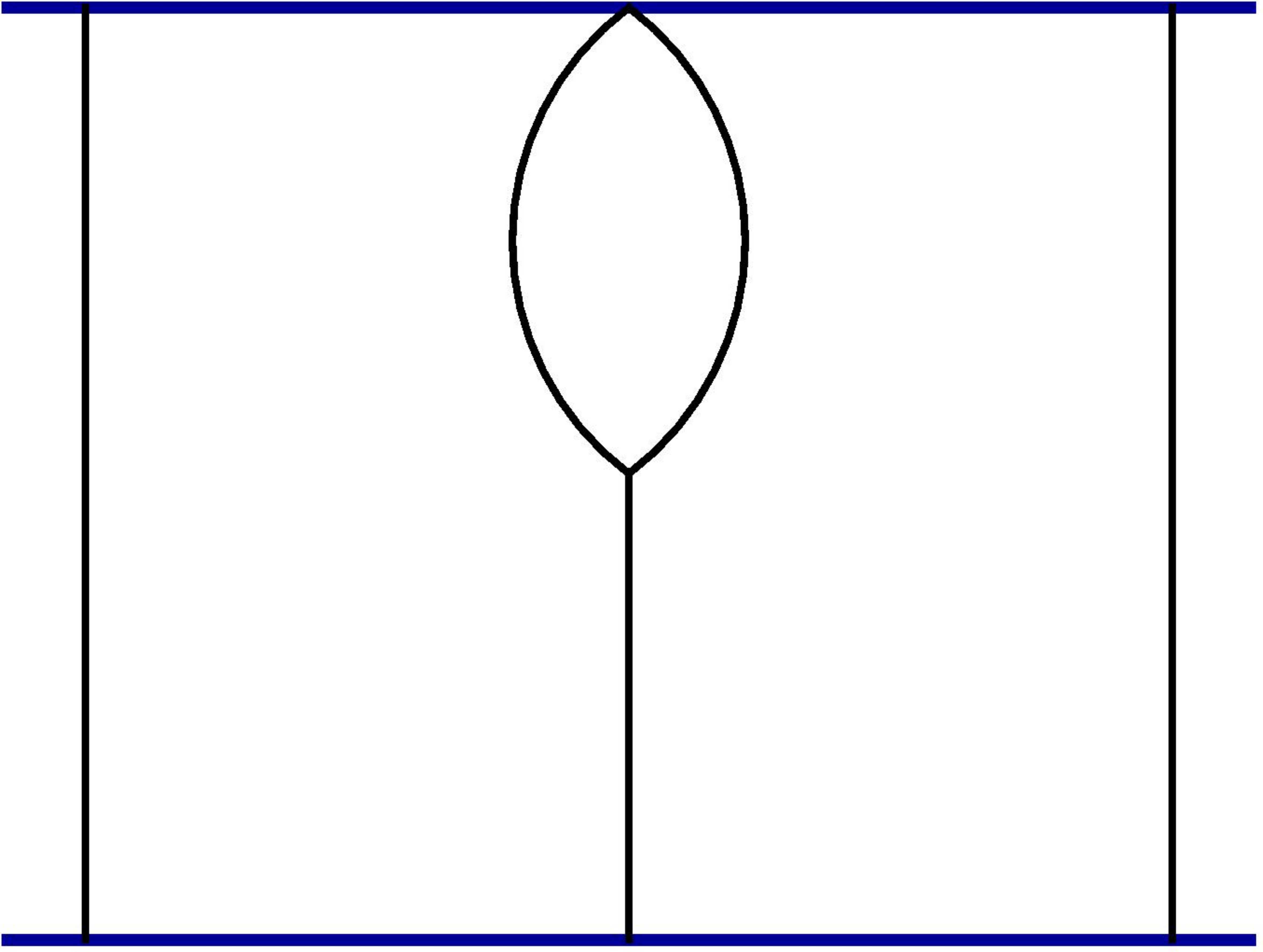}-\pictext{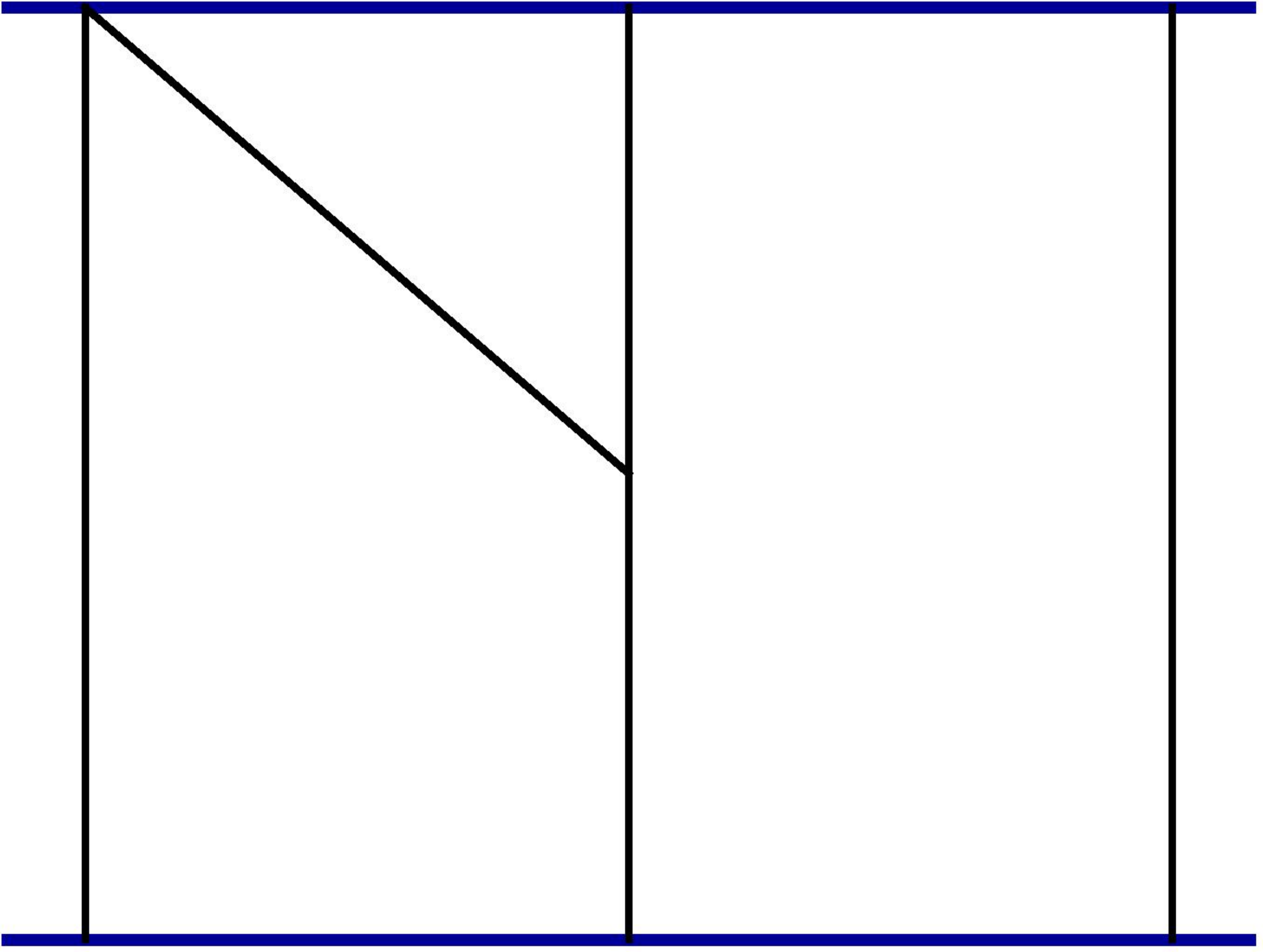}-\pictext{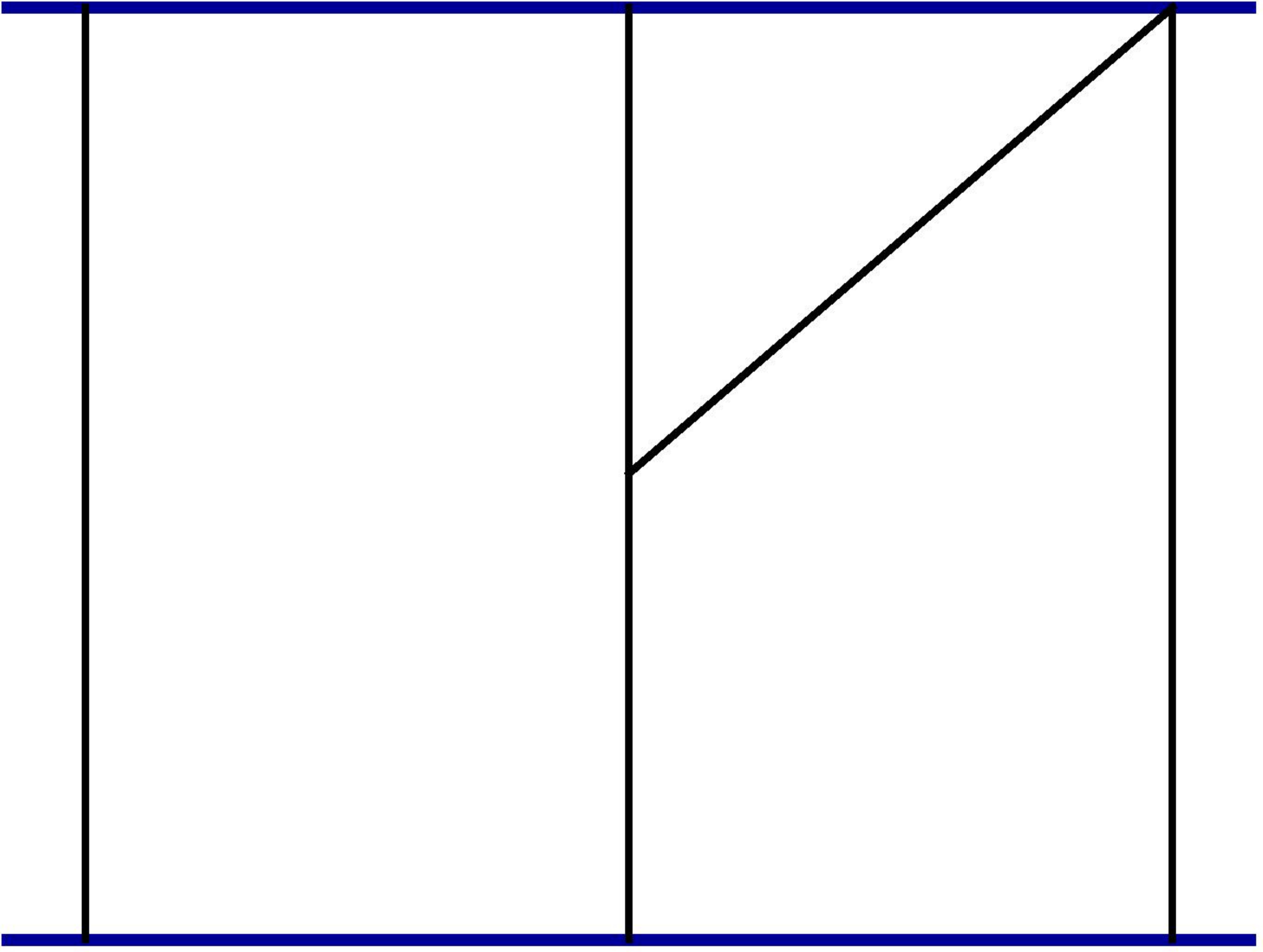}
\end{equation}
The last two diagrams cancel off the four middle terms in (\ref{Passarino-Veltman}). As shown in the appendix~\ref{self}, the bubble diagram cancels with the self-energy corrections in fig~\ref{fig1-allD}b. We are thus left with a single contribution, given by the second derivative
of the scalar H-graph (fig.~\ref{HD}b): 
\begin{equation}\label{K1}
 K_1(u,v;x)={\hat{\lambda }^2}\,\Delta H(x_1,\ldots, x_4 ),
\end{equation}
where the Laplacian is defined as
\begin{equation}
  \Delta =\left(\partial _1+\partial _2\right)^2.
\end{equation}
and
\begin{equation}\label{Hdef}
H(x_1, x_2, x_3, x_4)={1\over (4\pi^2)^5}\int\;\frac{d^4 z\;d^4 y}{(z-x_1)^2\,(z-x_4)^2\,(z-y)^2\,(y-x_2)^2\,(y-x_3)^2}
\end{equation}
The diagram is taken in the following kinematics:
\begin{equation}\label{kinematic}
 x_1=(u,0,0,1),~x_2=(0,0,0,1),~x_3=(x,0,0,0),~x_4=(x+v,0,0,0).
\end{equation}

We do not know if one can express the H-diagrams in terms of known functions. This is a difficult double-triangle two-loop integral, for which we were unable to
find any closed analytic expression. The best we could do is
a double-integral representation, presented in appendix~\ref{H-app}. However, for analyzing the strong and weak coupling limits it will suffice to know the asymptotic expressions of the H-diagram, which are relatively easy to get from the Feynman integral directly.

The problem of finding the Coulomb charge at the NLO boils down to solving the linear problem (\ref{eigenvalue}) with the kernel given by the sum of (\ref{Ktree}) and (\ref{K1}). Doing this exactly is rather difficult, as the Schr\"odinger operator becomes integro-differential, and the eigenvalue $\alpha $ enters the equation non-linearly. However, $\lambda $ is small and we can apply the usual formulas of quantum-mechanical perturbation theory. But as we do not know the leading-order wavefunctions explicitly, we will restrict ourselves to the two limiting cases of small $\hat{\lambda }$ and large $\hat{\lambda }$.

\section{Strong coupling}

At large $\hat{\lambda }$ the problem significantly simplifies, because then $\alpha $ is large, and the integral transform in (\ref{eigenvalue}) is highly peaked at small $u$ and $v$, because of the exponential suppression factor. In particular, the scale of variation of the wavefunction is proportional to $\hat{\lambda }^{-1/4}$, cf.~(\ref{spilarge}), and is thus much larger than the scale at which the integrand gets exponentially suppressed, which is $\alpha ^{-1}\sim \hat{\lambda }^{-1/2}$. We can thus neglect the shift in the argument of $\psi $ by $v-u$ and reduce the problem to the Schr\"odinger equation with a local effective potential
\begin{equation}\label{Veff}
 V_{\rm eff}(x)=-\int_{0}^{\infty }du\int_{0}^{\infty }dv \,
 \,{\rm e}\,^{-\frac{\alpha }{2}\left(u+v \right)}
 K(u,v ;x),
\end{equation}
Moreover, the kernel itself changes on the distances of order one, and we thus need to know its expansion at small  $u$ and $v$. This corresponds to the regime when the end-points of the H-diagram collide pairwise (fig.~\ref{BSEfig}). 
 The relevant limit  is computed in eq.~(\ref{H-short-dist}), with logarithmic accuracy:
\begin{equation}\label{Kappr}
 K_1(u,v;x)\simeq \frac{\hat{\lambda }^2\left(u^2\ln u ^2+v ^2\ln v^2\right)}{256\pi ^6\left(x^2+1\right)^3}\,.
\end{equation}
Substituting this result into (\ref{Veff}), we get for the effective potential:
\begin{equation}
 V_{\rm eff}(x)=-\frac{\hat{\lambda }}{4\pi ^2\left(x^2+1\right)}
 +\frac{\lambda \hat{\lambda }^2\ln\alpha }{2\pi ^6\alpha ^4\left(x^2+1\right)^3}.
\end{equation}

At strong coupling, when the eigenvalue is given just by the depth of the potential well, we get:
\begin{equation}
 \alpha =\frac{\sqrt{\hat{\lambda }}}{\pi }\left(1-\frac{\lambda \ln\hat{\lambda }}{2\hat{\lambda }}\right).
\end{equation}
The second term in the brackets is the correction to the ladder approximation. Since we have not determined normalization of the logs in (\ref{Kappr}), the answer is calculated with the logarithmic accuracy. It should be possible to more accurately determine all the constants and calculate the non-logarithmic term of order $\lambda /\hat{\lambda }$, but we will not attempt to do it here.

It is a non-trivial fact that the corrected Coulomb charge, as a function of $\lambda $ and $\vartheta$,  is still proportional to $\lambda ^{1/2}$ and thus can be directly compared to the string calculation at strong coupling. Provided that $\ln\hat{\lambda }$ can be traded for $\vartheta$,  in the strong-coupling limit, we get:
\begin{equation}\label{strong}
 \alpha =\frac{\sqrt{\lambda }}{2\pi }\,\,{\rm e}\,^{\frac{\vartheta}{2}}
 \left(1-2\vartheta\,{\rm e}\,^{-\vartheta}\right).
\end{equation}

\subsection{Comparison to string theory}

In string theory, the Wilson loop vev is given by the area of the minimal surface bounded by anti-parallel lines on the boundary of $AdS_5$, and interpolating between two points on $S^5$ with angular separation $\theta $. The solution for the minimal surface was found in \cite{Maldacena:1998im}. The area can be written in the parametric form\footnote{The same result can be extracted from the minimal surface bounded by a cusp \cite{Drukker:1999zq,Drukker:2011za}. We follow the notations and conventions of \cite{Drukker:2011za}.}:
\begin{eqnarray}\label{alphastring}
 \alpha &=&\frac{2\sqrt{\lambda }\left[E-\left(1-k^2\right)K\right]^2}{\pi k\sqrt{1-k^2}}
 \\ \label{thetastring}
 \vartheta&=&2\sqrt{2k^2-1}\,K,
\end{eqnarray}
where $K\equiv K(k^2)$ and $E=E(k^2)$ are the complete elliptic integrals of the first and second kind. The angle $\theta $ is real when the elliptic modulus satisfies $k^2<1/2$. Analytic continuation to imaginary $\theta $ corresponds to real $k^2>1/2$.

We are interested in the limit $\vartheta\rightarrow \infty $, which corresponds to  taking $k^2\rightarrow 1$. 
Expanding   (\ref{alphastring}), (\ref{thetastring}) in $1-k^2$ with the logarithmic accuracy, we get:
\begin{eqnarray}
 \alpha& =&\frac{2\sqrt{\lambda }}{\pi \sqrt{1-k^2}}\left(1+\frac{1-k^2}{2}\,
 \ln\left(1-k^2\right)+O\left(1-k^2\right)\right)
 \\
 \vartheta&=&-\ln\left(1-k^2\right)+\ln 16 +\frac{3\left(1-k^2\right)}{4}\,\ln\left(1-k^2\right)+O\left(1-k^2\right).
\end{eqnarray}
Consequently,
\begin{equation}
 \alpha =\frac{\sqrt{\lambda }}{2\pi }\,\,{\rm e}\,^{\frac{\vartheta}{2}}
 \left(1-2\vartheta\,{\rm e}\,^{-\vartheta}\right).
\end{equation}
 in an agreement with (\ref{strong}). 
 
 As emphasized in \cite{Correa:2012nk}, taking $\lambda $ large and then sending $\vartheta$ to  infinity is not quite the same as keeping $\lambda $ small and sending $\hat{\lambda }$ to infinity. We see however that potential order-of-limits ambiguities do not affect the agreement of the ladder resummation with string theory even at the NLO order.

\section{Weak coupling}

The first order of perturbation theory in $\lambda$ for the eigenvalue equation (\ref{eigenvalue}) produces the following answer for the correction to the Coulomb charge:
\bea\label{corr}
\delta\alpha={2\lambda\over \alpha_0}\,\int\limits_{-\infty}^{\infty} dx\,\int\limits_0^\infty du\,\int_{0}^{\infty }dv\,\,{\rm e}\,^{-{\alpha_0 \over 2}(u+v)} \psi_0(x)\,\psi_0(x-u+v)\,K_1(u, v, x),
\eea
where  $K_1$ is  first correction to the kernel from (\ref{K1})-(\ref{kinematic}).  At small $\hat{\lambda }$ the unperturbed wavefunction is given by 
\begin{equation}
 \psi _0(x)\simeq\left( \frac{\alpha_0 }{2}\right)^{\frac{1}{2}}\,{\rm e}\,^{-\frac{\alpha_0 }{2}|x|}\,\qquad (\hat{\lambda }\rightarrow 0).
\end{equation}
Since $\alpha _0$, given by  eq.~(\ref{weakalpha0}), is parametrically small in $\hat{\lambda }$, all exponential factors in (\ref{corr}) decay very slowly and to the first approximation can be set to one, which leads to 
\begin{equation}\label{simplified3loop}
 \delta \alpha =\lambda \hat{\lambda }^2
 \int\limits_{-\infty}^{\infty} dx\,\int\limits_0^\infty du\,\int_{0}^{\infty }dv\,
 \Delta H(x_1,\ldots ,x_4),
\end{equation}
where $x_i$ are expressed through  $u$, $v$ and $x$ according to (\ref{kinematic}), or according to fig.~\ref{BSEfig}. We  got the expected result, that to the leading order in $\hat{\lambda }$ it is enough to integrate the H-diagram along the Wilson loop. There are no ladders to sum, as extra rungs of the ladder are accompanied by extra powers of $\hat{\lambda }$. However, as we shall demonstrate in a moment, the integrals in (\ref{simplified3loop}) actually diverge. To understand how the divergence is cut off,  we need to go one step back to the all-loop expression (\ref{eigenvalue}). The exponential suppression factors in the exact formula becomes important on the scale
\begin{equation}\label{tIR}
 t_{\rm IR}\sim \alpha_0^{-1} \sim \hat{\lambda }^{-1}\,.
\end{equation}
This scale imposes an effective IR cutoff on the integrals in (\ref{simplified3loop}).  

\begin{figure}[t]
\begin{center}
 \centerline{\includegraphics[width=8cm]{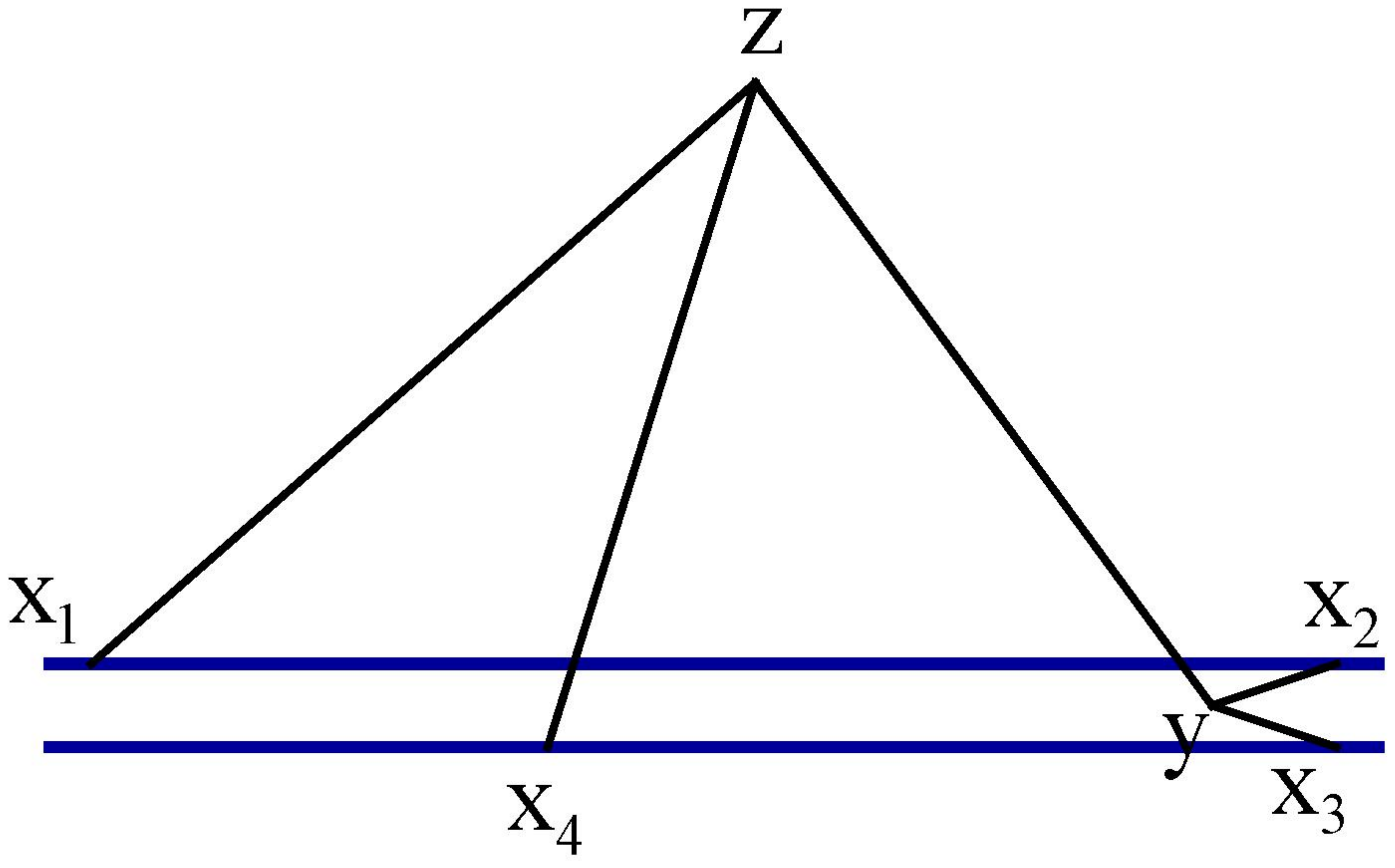}}
\caption{\label{danger}\small The dangerous region of integration in the H-diagram. }
\end{center}
\end{figure}
The integral (\ref{simplified3loop}) is convergent by power counting, which indicates that nothing special happens when all points  go to infinity simultaneously. The divergence comes from the configuration in which $x_2$ and $x_3$ are a finite distance from  one another, while $x_1$, $x_4$ go to infinity (fig.~\ref{danger}), and from a symmetric configuration in which $x_1$, $x_4$ stay close, while $x_2$, $x_3$ go to infinity. In this regime we can use the long-distance approximation (\ref{Long-Distance}) for the H-diagram. Moreover, we can apply the Laplacian only to the logarithmic prefactor, since the derivatives acting on the other terms in (\ref{Long-Distance}) raise powers in the denominator, and produce IR convergent integrals. Then, using $\Delta \ln x_{23}^2=4/x_{23}^2=4/(x^2+1)$, we can easily integrate over $x$:
\begin{equation}
 \int_{-\infty }^{+\infty }\frac{dx}{x^2+1}=\pi .
\end{equation}

We are thus left with the $u$, $v$ and $z$ integration:
\begin{equation}
 \delta \alpha =-2\,\frac{\lambda \hat{\lambda }^2}{4^4\pi ^7}
 \int\limits_0^\infty du\,\int_{0}^{\infty }dv\,
 \int_{}^{}\frac{d^4z}{z^2\left[\mathbf{z}^2+\left(u-z^0\right)^2\right]
 \left[\mathbf{z}^2+\left(v-z^0\right)^2\right]}\,.
\end{equation}
The factor of two takes into account the identical contribution from $x_{14}\sim 1$.
Doing the $u$ and $v$ integration first, we arrive at
\begin{equation}
 \delta \alpha =-\frac{\lambda \hat{\lambda }^2}{(2\pi) ^7}
 \int_{}^{}\frac{d^4z}{z^2\mathbf{z}^2}\,\left(\frac{\pi }{2}+\arctan\frac{z^0}{|\mathbf{z}|}\right)^2.
\end{equation}
The integral is most easily computed by changing variables to $\rho=\sqrt{z^2}$ and $\varphi =\arctan(z^0/|\mathbf{z}|)$:
\begin{equation}
 \delta \alpha =-\frac{\lambda \hat{\lambda }^2}{32\pi ^6}
 \int_{}^{t_{IR}}\frac{d\rho }{\rho }\,
 \int_{-\frac{\pi }{2}}^{+\frac{\pi }{2}}\left(\frac{\pi }{2}+\varphi \right)^2.
\end{equation}
The logarithmic divergence at small $\rho $ is fake. It arises because we have initially assumed that $\rho \gg  1$, and neglected, for example, the separation between the two sides of the Wilson loop. The neglected contributions regularize the divergence at some $\rho \sim 1$. On the contrary the IR divergence at large $\rho $ is real, and is regularized only by the resummation of an infinite number of ladder diagrams. The resulting cutoff, up to a numerical factor, is given by (\ref{tIR}). Taking this relationship into account, we get the final answer for the correction to the Coulomb charge (\ref{weakalpha0}):
\bea\label{da}
\delta \alpha=\frac{1}{96 \pi^3}\,\lambda\,\hat{\lambda}^2\,\log{\hat{\lambda}}
\eea

Taken at face value, the result (\ref{da}) is a three-loop single-logarithmic correction, which combines with the single-log terms in (\ref{weakalpha0}). There is also a finite $O(\lambda \hat{\lambda }^2)$ term, that we have not calculated. Clearly, the higher orders in the ladder expansion will produce single-log contributions of the form  $\lambda ^2\hat{\lambda }\ln \hat{\lambda }$ and $\lambda ^3\ln\hat{\lambda }$, which originate, for instance, from the purely gluonic H-diagram. It would be interesting to compute these log-enhanced contributions, which complete the perturbative three-loop calculation of the Coulomb charge in the NLLO.

\section{Conclusions}

The leading-order ladder resummation can be done also for the Wilson loops on $S^3\times R^1$ \cite{Correa:2012nk}. In $R^4$ this configuration maps to a cusp with the opening angle $\varphi $, where $\varphi $ is the angular separation of the Wilson lines on the sphere \cite{Correa:2012at}. Our calculation corresponds to the limiting case  $\varphi \rightarrow \pi$. It would be interesting to evaluate the NLO correction for general $\varphi $, especially since the  Schr\"odinger problem is exactly solvable at $\varphi =0$. This latter case corresponds to the straight Wilson line, with  the scalar coupling jumping from $0$ to $\cos\theta $ at some point on the contour.

It would be also interesting to reproduce the ladder limit (\ref{CHMSlimit}) from the TBA equations, that in principle describe the Coulomb charge at any value of $\theta $ and $\lambda $ \cite{Drukker:2012de,Correa:2012hh}. The TBA equations are highly non-linear, and in general are difficult to solve, but when $\vartheta\rightarrow \infty $ they should reduce to a simple Schr\"odinger equation (\ref{Sch}). Such a dramatic simplification is not unconceivable, as $\vartheta$ plays the r\^ole of a chemical potential in TBA. Large real chemical potential (real $\theta $ actually corresponds to an imaginary chemical potential) should  strongly emphasize the degrees of freedom with positive charge, which perhaps can be taken into account by rearranging the ground state of the auxiliary spin system. Whatever the underlying reason for the drastic reduction in complexity in comparison to the general case, in this corner of the parameter space the TBA equations should be solvable.

\subsection*{Acknowledgments}
We would like to thank N.~Drukker for many useful discussions.
K.Z. would like to thank IAS at the Hebrew University of Jerusalem and the Perimeter Institute for kind hospitality during the course of this work.
The work of K.Z. was supported in part by  the RFFI grant 10-02-01315, and in part
by the Ministry of Education and Science of the Russian Federation
under contract 14.740.11.0347. The work of D.B. was supported in part by grants RFBR 11-01-00296-a, 11-01-12037-ofi-m-2011 and in part by grant for the Support of Leading Scientific Schools of Russia NSh-4612.2012.1.

\appendix

\section{Cancellation of the self-energy corrections}
\label{self}

Here we will show that the one-loop correction to the propagator (fig.~\ref{fig1-allD}b) exactly cancels with the bubble graph in the right-hand-side of eq.~(\ref{Y's}). Both of these diagrams diverges, and it is necessary to regularize them exactly in the same way, to make sure that no finite leftover survives the cancellation. We use the usual dimensional reduction scheme.

In the dimensional reduction regularization, the one-loop scalar propagator is given by \cite{Erickson:2000af}:
\begin{equation}\label{corrprop}
 D^{\rm 1-loop}(r)=\frac{\Gamma \left(\frac{D}{2}-1\right)}{4\pi ^{\frac{D}{2}}r^{D-2}}
 -\frac{\lambda \Gamma ^2\left(\frac{D}{2}-1\right)}{16\pi ^D\left(D-3\right)
 \left(4-D\right)r^{2D-6}}\,.
\end{equation}

The bubble diagram then gives:
\begin{equation}
 D^{\rm bubble}(r)=\frac{2\lambda \Gamma ^3\left(\frac{D}{2}-1\right)}{\left(4\pi ^{\frac{D}{2}}\right)^3}
 \int_{}^{}\frac{d^Dz}{z^{2D-2}(r-z)^{D-2}}
 =
 \frac{\lambda \Gamma ^2\left(\frac{D}{2}-1\right)}{16\pi ^D\left(D-3\right)
 \left(4-D\right)r^{2D-6}}\,,
\end{equation}
which exactly cancels the second term in (\ref{corrprop}), even before taking the $D\rightarrow 4$ limit.

\section{H-diagram}\label{H-app}

The following auxiliary integral is useful:
\begin{equation}
 \int_{}^{}\frac{d^4z\,d^4y}{\left(z^2+\mu ^2\right)^2\left(y^2+\nu ^2\right)^2\left(z-y+a\right)^2}
 =\frac{\pi ^4}{4a^2}\left(\ln^2\frac{\omega _+}{\omega _-}-\ln^2\frac{\mu ^2}{\nu ^2}\right),
\end{equation}
where $\omega _\pm$ are the roots of the quadratic equation
\begin{equation}
 \omega _\pm^2-\left(\mu ^2+\nu ^2+a^2\right)\omega _\pm+\mu ^2\nu ^2=0.
\end{equation}
With the help of this formula, we can get for the H-diagram a double-integral representation over two sets of Feynman parameters:
\begin{equation}
 H=\frac{1}{\left(4\pi \right)^6}
 \int_{0}^{1}d\kappa _1\int_{0}^{1}d\kappa _2\,\,
 \frac{1}{a^2}\left[\ln^2\frac{\omega _+}{\omega _-}-\ln^2\frac{\kappa _1\left(1-\kappa _1\right)x_{14}^2}{\kappa _2\left(1-\kappa _2\right)x_{23}^2}\right],
\end{equation}
where now $\omega _\pm$ are the roots of
\begin{eqnarray}
 &&\omega _\pm^2-\left[
 \kappa _1\left(1-\kappa _2\right)x_{13}^2+\kappa _2\left(1-\kappa _1\right)x_{24}^2+\kappa _1\kappa _2x_{12}^2+\left(1-\kappa _1\right)\left(1-\kappa _2\right)x_{34}^2
 \right]\omega _\pm
\nonumber \\ 
 &&+\kappa _1\kappa _2\left(1-\kappa _1\right)\left(1-\kappa _2\right)x_{14}^2x_{23}^2=0,
\end{eqnarray}
and
\begin{equation}
 a=\kappa _1x_{14}-\kappa _2x_{23}-x_{34}.
\end{equation}
It is also true that
\begin{eqnarray}
 a^2&=& \kappa _1\left(1-\kappa _2\right)x_{13}^2+\kappa _2\left(1-\kappa _1\right)x_{24}^2+\kappa _1\kappa _2x_{12}^2+\left(1-\kappa _1\right)\left(1-\kappa _2\right)x_{34}^2
\nonumber \\
 &&-\kappa _1\left(1-\kappa _1\right)x_{14}^2-\kappa _2\left(1-\kappa _2\right)x_{23}^2.
\end{eqnarray}

\subsection{Short-distance asymptotics}

First we consider the case when $|x_{12}|\sim |x_{34}|\ll |x_{23}|$.
The H-diagram is non-singular in the limit $x_{12},x_{34}\rightarrow 0$, and on dimensional grounds we expect the leading term of the form $\,{\rm const}\,/x_{23}^2$. As $1/x_{23}^2$ is annihilated by the Laplacian at $x_2\neq x_3$, we will need the next term in the expansion in $x_{12}$, $x_{34}$. This term appears at the quadratic order and contains an extra logarithmic enhancement. The large logarithm $\ln(x_{12}^2/x_{23}^2)$ arises from the region of integration where both integration points simultaneously approach $x_{1,2}$ or $x_{3,4}$. When both $z$ and $y$ are close to $x_1$, $x_2$, each of the two "long" propagators can then be approximated by $1/x_{23}^2$, and we are left with the integral
\begin{equation}
 \frac{1}{\left(4\pi^2 x_{23}^2 \right)^2}\int_{}^{}\frac{d^4zd^4y}{\left(4\pi ^2\right)^3\left(x_1-z\right)^2\left(z-y\right)^2\left(y-x_2\right)^2}+\left(1,2\rightarrow 3,4\right).
\end{equation}
The integral is the Fourier transform of $1/p^6$, equal to $x_{12} ^2\ln x_{12
} ^2/2^7\pi ^2$. Taking into account the contribution from the other coincidence limit, we obtain:
\begin{equation}\label{H-short-dist}
 H=\frac{\,{\rm const}\,}{x_{23}^2}+\frac{x_{12} ^2\ln\frac{x_{12} ^2}{x_{23}^2}+x_{34}^2\ln\frac{x_{34} ^2}{x_{23}^2}}{2^{11}\pi ^6x_{23}^4}+\ldots .
\end{equation}

\subsection{Long-distance asymptotics}

Next we consider the case when $|x_{23}|\ll |x_{12}|\sim |x_{34}|$ (fig.~\ref{danger}). When $x_{23}\rightarrow 0$,  the $y$-integration in (\ref{Hdef}) is log-divergent at $y\rightarrow x_{2,3}$. The divergence is cut off at $y\sim x_{23}$, and we can thus replace
\begin{equation}\label{Long-Distance}
 H\simeq -\frac{\ln x_{23}^2}{4^5\pi ^8}
 \int_{}^{}\frac{d^4z}{\left(z-x_1\right)^2\left(z-x_2\right)^2\left(z-x_3\right)^2}\,.
\end{equation}

\bibliographystyle{nb}

\end{document}